\documentclass[aps,prb,twocolumn,showpacs,amsfonts]{revtex4}
\usepackage{graphicx}
\usepackage{dcolumn}
\usepackage{bm}

\begin{document}

\title{Formation of density singularities in ideal hydrodynamics of freely cooling
inelastic gases: a family of exact solutions}

\author{Itzhak Fouxon}
\author{Baruch Meerson}
\author{Michael Assaf}
\author{Eli Livne}

\affiliation{Racah Institute of Physics, Hebrew University of Jerusalem,
Jerusalem 91904, Israel}
\date{\today }

\begin{abstract}

We employ granular hydrodynamics to investigate a paradigmatic problem of
clustering of particles in a freely cooling dilute granular gas. We consider
large-scale hydrodynamic motions where the viscosity and heat conduction can be
neglected, and one arrives at the equations of ideal gas dynamics with an
additional term describing bulk energy losses due to inelastic collisions. We
employ Lagrangian coordinates and derive a broad family of exact non-stationary
analytical solutions  that depend only on one spatial coordinate. These
solutions exhibit a new type of singularity, where the gas density blows up in a
finite time when starting from smooth initial conditions. The density blowups
signal formation of close-packed clusters of particles. As the density blow-up
time $t_c$ is approached, the maximum density exhibits a power law $\sim
(t_c-t)^{-2}$. The velocity gradient blows up as $\sim - \,(t_c-t)^{-1}$ while
the velocity itself remains continuous and develops a cusp (rather than a shock
discontinuity) at the singularity. The gas temperature vanishes at the
singularity, and the singularity follows the isobaric scenario: the gas pressure
remains finite and approximately uniform in space and constant in time close to
the singularity. An additional exact solution shows that the density blowup, of
the same type, may coexist with an ``ordinary" shock, at which the hydrodynamic
fields are discontinuous but finite. We confirm stability of the exact solutions
with respect to small one-dimensional perturbations by solving the ideal
hydrodynamic equations numerically. Furthermore, numerical solutions show that
the local features of the density blowup hold universally, independently of
details of the initial and boundary conditions.

\end{abstract}


\maketitle

\section{Introduction}

Structure formation in many-body systems is one of central problems of
non-equilibrium physics. A most spectacular phenomena of structure formation is
clustering of matter. Here an initially structureless, almost homogeneous
distribution of particles of matter self-assembles into clusters. Such an
evolution cannot proceed indefinitely in a gas where interactions between the
particles are (i) short-range  and (ii) Hamiltonian. There are two important
classes of gases where one of these two properties is violated, and clustering
can occur. In the first one \textit{long-range} forces, such as gravity, are
present. Clustering provides here a natural mechanism of star formation
\cite{star} and of the large-scale structure of the Universe \cite{universe}.
The second class of systems are dissipative systems where interactions between
the particles, at the level of an effective description, are
\textit{non-Hamiltonian}. One well-known example is optically thin gases and
plasmas that cool by their own radiation, and dense condensations develop
\cite{Parker,Field,meersonRMP}. In this paper we consider a more basic
non-Hamiltonian many-body system: a gas of inelastically colliding macroscopic
particles, or granular gas. Here particles lose energy to their internal degrees
of freedom.

The granular gas is the low-density limit of granular flows
\cite{BP,Goldhirsch2}. In its simplest version, the granular gas model deals
with a dilute assembly of identical hard spheres (with diameter $\sigma$ and
unit mass) who lose energy at instantaneous binary collisions in such a way that
the normal component of the relative velocity of particles is reduced by a
constant factor $0\leq r<1$ (the coefficient of normal restitution) upon each
collision. Granular gases exhibit a plethora of structure forming instabilities,
including the clustering instability of a freely cooling homogeneous inelastic
gas \cite{Hopkins,Goldhirsch,McNamara1,McNamara2,Ernst,Brey,Luding,van
Noije,Ben-Naim2,ELM,MP,Garzo}. This instability brings about the generation of a
macroscopic flow and formation of dense clusters of particles.

The natural language for a theoretical description of macroscopic flows of a
granular gas is the Navier-Stokes granular hydrodynamics \cite{BP,Goldhirsch2}.
Although the criteria of its validity are quite restrictive, see below, granular
hydrodynamics is instrumental for theoretical investigations, and often
prediction, of a host of collective effects in granular flows. Recently,
applications of granular hydrodynamics have been extended to non-stationary
flows of granular gases \cite{ELM,Bromberg,Volfson,Fouxon1}. The non-stationary
settings provide sharp tests to continuum models of granular flows and help
evaluate their domains of validity. This is especially true when the
time-dependent solutions of the continuum equations develop finite-time
singularities \cite{Kadanoff}, such as the recently predicted density blowup in
freely cooling granular gases: at zero gravity \cite{ELM,Fouxon1}, and at finite
gravity \cite{Volfson}.

We will assume throughout this paper nearly elastic particle collisions,  a very
small gas density (that we denote by $\rho$), and a very small Knudsen number:
\begin{equation}\label{threeineq}
    1-r\ll 1\,,\;\;\;\;\;\rho \sigma^d\ll 1\,,\;\;\;\;
    \mbox{and}\;\;\;\;\;l_{free}/L\ll1\,.
\end{equation}
Here $d>1$ is the dimension of space, $l_{free}$ is the mean free path of the
particles, and $L$ is the characteristic length scale of the hydrodynamic
fields. Under these assumptions the Navier-Stokes hydrodynamics provides a
quantitatively accurate leading-order theory for granular gases
\cite{BP,Goldhirsch2}. Previous works \cite{Goldhirsch,McNamara1} employed
hydrodynamic equations to show that, for sufficiently large systems, the
homogeneously cooling state of the granular gas is linearly unstable. The
unstable perturbations grow via two (linearly) independent modes: the shear mode
corresponding to the development of a macroscopic solenoidal flow, and the
clustering mode corresponding to the development of a potential flow that causes
formation of clusters of particles. How does the clustering mode develop beyond
the linear regime, and do the hydrodynamic nonlinearities arrest the density
growth? For the clustering instability in large systems (as well as for the
gravitational instability \cite{universe}) the growing perturbations bring the
system into a fully developed non-linear regime
\cite{Goldhirsch,McNamara2,Luding,Ben-Naim2,ELM,MP}. Therefore, one has to deal
with fully non-linear hydrodynamic equations. Not restricting ourselves to a
close proximity to the homogeneously cooling state, we can formulate the problem
in a more general way and explore different nonlinear flows of a freely evolving
granular gas.

Solving the hydrodynamic equations analytically, and even numerically, is a
difficult task, and additional simplifications are needed. We will make
\textit{two} additional simplifying assumptions in this work. First, assuming a
large-scale flow, we will neglect the viscous and heat conduction terms in the
hydrodynamic equations. Second, we will assume that the macroscopic flow is
one-dimensional. A natural environment for the second assumption is provided by
the geometry of a narrow channel with perfectly elastic side walls that we adopt
here, following Refs. \cite{ELM,MP}. Although the microscopic motion of
particles in the channel remains two- or three-dimensional (2d or 3d), the
macroscopic flow depends only on one spatial coordinate: the coordinate along
the channel. This is because, in a narrow channel, both the shear mode and the
clustering mode in the transverse directions are suppressed (see Refs.
\onlinecite{ELM,MP} for detail).  As a result, one can focus on the development
of the (one-dimensional) clustering mode as it enters a strongly nonlinear
regime.

Working in the channel geometry, Efrati \textit{et al.} \cite{ELM} considered
the long-wavelength limit of the clustering instability, when the linear growth
rate of the instability is the highest. In this case the inelastic energy loss
of the gas is the fastest process, and the gas pressure drops, almost
instantaneously, to a very small value. The further dynamics are then (almost)
purely inertial which would bring about a finite-time blow-up of the velocity
gradient and, therefore, of the density. This is the well known blow-up of the
free flow \cite{Whitham}. Its signatures were observed in the numerical solution
of the hydrodynamic equations \cite{ELM} until the maximum gas density became so
high that the numerical scheme lost accuracy. The numerical results of Ref.
\cite{ELM} were tested in molecular dynamics (MD) simulations in 2d of a freely
cooling gas of inelastically colliding disks in a long and narrow channel
\cite{MP}. The MD simulations supported the free-flow blow-up scenario until the
time when the gas density approached the hexagonal close-packing value, and the
further density growth was arrested.

As we have recently found \cite{Fouxon1}, the free flow asymptotics does not
hold all the way to the density blowup. Very close, in time and in space, to the
(attempted) free-flow singularity, the compressional heating starts to act and
temporarily stabilizes the gas temperature at a small but finite value. As a
result, the gas pressure again becomes important: it breaks the purely inertial
dynamics and, though unable to stop the density blowup, dramatically changes the
local blowup properties. It turns out that, after a brief crossover,  the
further dynamics obey a new blowup scenario, not limited to the long-wavelength
limit \cite{Fouxon1}. The new scenario has the following features. As the
blow-up time $t_c$ is approached, the maximum density exhibits a power law $\sim
(t_c-t)^{-2}$. The velocity gradient blows up as $\sim  - (t_c-t)^{-1}$, whereas
the velocity itself remains continuous and develops a cusp, rather than a shock
discontinuity, at the singularity. The gas temperature vanishes at the
singularity, but the pressure there remains finite. This blowup, which obeys the
above asymptotic laws near the singularity, emerges universally, that is for
generic initial and boundary conditions.  (Note that, in the long-wavelength
limit, the crossover from the free flow regime to the finite-pressure regime
does not happen when the initial gas density is not very small. In this case
excluded particle volume effects interfere in the dynamics, and arrest the
density growth, \textit{before} the crossover has a chance to occur, as indeed
was observed in Ref.~\cite{MP}.)

These findings are based on extensive numerical simulations (numerical solutions
of the hydrodynamic equations for a host of initial and boundary conditions) and
a family of exact analytical solutions of the hydrodynamic equations, without
and with shocks, briefly announced in Ref. \cite{Fouxon1}. In the present work
we give a detailed account of the exact solutions and investigate their
structure close to the singularity. We verify the exact solutions numerically.
We show that the singularity follows the isobaric scenario: the gas pressure is
approximately uniform in space and constant in time in a close vicinity of the
developing singularity. Finally, we evaluate the validity of the exact solutions
close to the singularity by estimating the role of additional physical
processes: the viscosity, heat conduction and excluded particle volume effects.

The remainder of the paper is organized as follows. In Section II we introduce
the equations of \textit{ideal} granular hydrodynamics (IGHD) and discuss their
general properties. In Section \ref{analytic} we employ Lagrangian coordinates
and derive a family of exact solutions of the IGHD equations, consider some
particular cases of the solutions and investigate global and local properties of
the solutions. In Section \ref{shocks} we adapt the exact solutions to describe
a different setting, where a piston moves into a granular gas at rest, and a
density blowup, developing at the piston, coexists with an ``ordinary" shock
wave propagating into the gas. Besides demonstrating the presence of the two
different types of singularities in the same system, this solution allows an
arbitrary initial density profile at large distances, showing that the density
blowup is a local process. Section \ref{numerics} presents the results of
numerical solutions of the IGHD equations that confirm stability of the exact
solutions with respect to small one-dimensional perturbations and establish
universality of the density blowup for different initial conditions. In Section
\ref{regularization} we discuss the role of non-ideal effects, neglected in our
solutions, close to the singularity. In Section \ref{summary} we summarize our
results and discuss their bearing on cluster formation.

\section{Ideal hydrodynamics of a freely cooling granular gas} \label{hydrodynamics}

Under the three strong inequalities~(\ref{threeineq}), the Navier-Stokes
granular hydrodynamics provides a quantitatively accurate leading-order theory
for granular gases. It deals with three coarse-grained fields: the mass density
$\rho(\mathbf{x}, t)$, the mean flow velocity ${\mathbf v}({\mathbf x}, t)$ and
the granular temperature $T({\mathbf x}, t)$. An additional coarse-grained
field, the pressure $p({\mathbf x}, t)$, is related to the density and
temperature by the perfect gas equation of state $p=\rho T$. Assuming a
one-dimensional macroscopic flow, we can write these equations as
\begin{eqnarray}&&
\frac{\partial \rho}{\partial t}+\frac{\partial(\rho v)}{\partial x}=0,
\label{hydrodynamics1}\\&& \rho\left(\frac{\partial v}{\partial t}+
v\frac{\partial v}{\partial x}\right)=-\frac{\partial (\rho T)}{\partial
x}+\nu_0\frac{\partial}{\partial x}\left(\sqrt{T}\frac{\partial v}{\partial
x}\right), \label{hydrodynamics2}
\\&& \frac{\partial T}{\partial t}+
v\frac{\partial T}{\partial x}=-(\gamma-1) T\frac{\partial v}{\partial
x}-\Lambda\rho
T^{3/2}\nonumber\\&&+\frac{\kappa_0}{\rho}\frac{\partial}{\partial x}
\left(\sqrt{T}\frac{\partial T}{\partial x}\right)
+\frac{\nu_0(\gamma-1)\sqrt{T}}{\rho}\left(\frac{\partial v}{\partial
x}\right)^2\,.\label{hydrodynamics3}
\end{eqnarray}
Here $\gamma$ is the adiabatic index of the gas ($\gamma=2$ and $5/3$ for $d=2$
and $d=3$, respectively), $\Lambda=2 \pi^{(d-1)/2} (1-r^2) \sigma^{d-1}/[d\,
\Gamma(d/2)]$ (see \textit{e.g.} \cite{Brey}), and $\Gamma(\dots)$ is the gamma
function. In addition, $\nu_0=(2\sigma\sqrt{\pi})^{-1}$ and $\kappa_0=4\nu_0$ in
2d, and $\nu_0=5(3\sigma^2 \sqrt{\pi})^{-1}$ and $\kappa_0=15\nu_0/8$ in 3d, see
Ref. \onlinecite{BP}. The only difference between
Eqs.~(\ref{hydrodynamics1})-(\ref{hydrodynamics3}) and the standard gas dynamic
equations for a dilute gas of \textit{elastically} colliding spheres is the
presence in Eq.~(\ref{hydrodynamics3}) of the inelastic energy loss term
$-\Lambda \rho T^{3/2}$.

There are three types of dissipative terms in
Eqs.~(\ref{hydrodynamics1})-(\ref{hydrodynamics3}): the viscous terms in
Eqs.~(\ref{hydrodynamics2}) and (\ref{hydrodynamics3}), the heat conduction term
in Eq.~(\ref{hydrodynamics3}) and the energy loss term in
Eq.~(\ref{hydrodynamics3}). The viscous and heat conduction terms include
spatial gradients of the hydrodynamic fields, whereas the energy loss term is
independent of the gradients. When the characteristic hydrodynamic length scale
of the flow is sufficiently large, the viscous and heat conduction terms can be
neglected, while the energy loss term should be kept, and we arrive at the
equations of \textit{ideal} granular hydrodynamics (IGHD):
\begin{eqnarray}&&
\frac{\partial \rho}{\partial t}+\frac{\partial(\rho v)}{\partial x}=0,\ \
\rho\left(\frac{\partial v}{\partial t}+v\frac{\partial v}{\partial x}\right)= -
\,\frac{\partial (\rho T)}{\partial x}, \label{a311} \\&& \frac{\partial
T}{\partial t} +v \frac{\partial T}{\partial x}=-(\gamma-1)T\frac{\partial
v}{\partial x}-\Lambda\rho T^{3/2}. \label{a322}
\end{eqnarray}
For consistency, all the assumptions must be checked \textit{a posteriori},
after a hydrodynamic problem in question is solved, and the hydrodynamic length
and time scales are found.

The basic state of the freely cooling gas is the homogeneously cooling state
described by the Haff law \cite{Haff}:
\begin{equation}
\rho= \rho_0,\;\;\;\; v= 0,\;\;\;\; T=\frac{T_0}{\left(1+\Lambda \rho_0
T_0^{1/2}t/2\right)^2}\,. \label{Haff}
\end{equation}
This state corresponds to the initial conditions $\rho(x,t=0)=\rho_0=const$,
$T(x,t=0)=T_0=const$ and $v(x,t)=0$. Obviously, the IGHD equations reproduce the
Haff's law exactly, as the homogeneously cooling state does not include
gradients of the hydrodynamic fields.

A more meaningful example of a situation where the IGHD applies is provided by
the linear stability analysis of the homogeneously cooling state. This analysis,
in the framework of the complete, non-ideal Navier-Stokes hydrodynamics
(\ref{hydrodynamics1})-(\ref{hydrodynamics3}) and its extensions was performed
by many workers, starting from Goldhirsch and collaborators \cite{Goldhirsch}
and McNamara \cite{McNamara1}. The main results of the linear stability analysis
can be described, at the level of order-of-magnitude estimates, as follows. The
evolution of a small sinusoidal perturbation with the wave number $k$ is
determined by two competing processes. The inelastic energy loss tends to
enhance the fluctuations on the cooling time scale $[(1-r^2)\sigma^{d-1}\rho
\sqrt{T}]^{-1}$ which is $k$-independent. In its turn, the viscosity and thermal
conduction tend to erase the perturbation on a time scale $[k^2 l_{free}^2
\rho\sigma^{d-1}\sqrt{T}]^{-1}$, where $l_{free}\sim 1/(n \sigma^{d-1})$ is the
mean free path and $l_{free}^2\rho\sigma^{d-1}\sqrt{T}$ is the characteristic
value of the viscosity/heat conductivity. Balancing these two time scales, one
obtains the critical wave number $k_c\sim l_{free}^{-1}\sqrt{1-r}$ so that the
perturbations with $k<k_c$ grow, while the short-wavelength perturbations,
$k>k_c$, decay \cite{Goldhirsch,McNamara1}. In a narrow channel, such that
perturbations in the transverse direction with wave numbers smaller than $k_c$
are not allowed, only perturbations along the channel will grow. As a result, by
the end of the linear stage, the hydrodynamic fields are effectively
one-dimensional and have characteristic length scales of the order of
$l_{free}/\sqrt{1-r}$ or longer. The validity of the Navier-Stokes hydrodynamics
(\ref{hydrodynamics1})-(\ref{hydrodynamics3}) for the description of the whole
range of unstable perturbations demands a strong inequality $\sqrt{1-r} \ll 1$.
On the other hand, the IGHD model~(\ref{a311}) and (\ref{a322}) is valid if we
demand a strong inequality $k\ll k_c$. Indeed, one can check that, in this case,
the growth rate of the clustering instability, as obtained from the IGHD,
coincides in the leading order in $k/k_c$ with that obtained from the full
hydrodynamic equations.

In Section \ref{regularization} we will perform a consistency check, and
establish the validity domain, of our exact \textit{nonlinear} solutions of the
IGHD equations (\ref{a311}) and~(\ref{a322}). Now let us discuss the basic
properties of these equations. Although much simpler than the non-ideal
equations (\ref{hydrodynamics1})-(\ref{hydrodynamics3}), the nonlinear IGHD
equations still present a hard mathematical problem. Going back to elastic
particle collisions, $\Lambda=0$, one recovers the ordinary ideal gas dynamics
in one dimension. Among most interesting solutions are those describing the
development of wave-breaking singularities when starting from smooth initial
data \cite{Landau,Chorin}. Note that, even for $\Lambda=0$, the general initial
value problem is not soluble analytically, except for the particular case of an
isentropic flow, where the entropy per unit volume $s(\rho,T)=\rho \ln
\left(T/\rho^{\gamma-1}\right)$ is uniform in space and constant in time
\cite{Landau,Chorin}. Needless to say, at $\Lambda> 0$, Eqs. ~(\ref{a311}) and
(\ref{a322}) do not allow isentropic solutions. The total entropy of the fluid
$S=\int s(\rho,T)\,dx$, governed by Eq. (\ref{a311}) and (\ref{a322}), is
monotone decreasing:
\begin{equation}
\frac{dS}{dt}=-\Lambda \int \rho^2 T^{1/2} dx\,, \label{entropy}
\end{equation}
where we have assumed that there is no net entropy flux from the boundaries. As
expected from the microscopic picture, the local entropy loss rate in
Eq.~(\ref{entropy}) is proportional to the particle collision rate. The entropy
loss implies that the system may exhibit self-organization phenomena
\cite{Haken} as is indeed observed in the process of clustering instability.

Before deriving a family of exact solutions exhibiting a finite-time density
blowup, we note two rescaling symmetries of Eqs.~(\ref{a311}) and (\ref{a322}).
The first symmetry relates solutions at different $\Lambda>0$. If $\rho(x, t)$,
$v(x, t)$ and $T(x, t)$ solve Eqs.~(\ref{a311}) and (\ref{a322}) at some
$\Lambda$, then the rescaled fields $\rho[(\Lambda'/\Lambda)x,
(\Lambda'/\Lambda)t]$, $v[(\Lambda'/\Lambda)x, (\Lambda'/\Lambda)t]$ and
$T[(\Lambda'/\Lambda)x, (\Lambda'/\Lambda)t]$ solve the same system with the
cooling coefficient $\Lambda'$. Therefore, the general mathematical properties
of Eqs.~(\ref{a311}) and (\ref{a322}), such as the existence of singularities,
are the same for any $\Lambda>0$. The second symmetry relates solutions of
Eqs.~(\ref{a311}) and (\ref{a322}) at the same $\Lambda$. If $\rho$, $v$ and $T$
solve Eqs. (\ref{a311}) and (\ref{a322}), then ${\tilde \rho}$, ${\tilde v}$ and
${\tilde T}$ defined by
\begin{eqnarray}&&
{\tilde \rho}(x, t)=\alpha \rho(\alpha x, \alpha\sqrt{\beta} t),\ \ {\tilde
v}(x, t)=\sqrt{\beta} v(\alpha x, \alpha\sqrt{\beta} t),\ \nonumber\\&& {\tilde
T}(x, t)= \beta T( \alpha x, \alpha\sqrt{\beta}t)\,, \label{symmetry}
\end{eqnarray}
with any $\alpha>0$ and $\beta>0$, also solve Eqs.~(\ref{a311}) and
(\ref{a322}). These symmetries are exploited in the following.

\section{Development of density singularities} \label{analytic}

\subsection{Lagrangian coordinates and exact solutions}
\label{Lagrangian}

Let us rewrite the governing equations (\ref{a311}) and (\ref{a322}) in terms of
the pressure $p=\rho T$, rather than temperature:
\begin{eqnarray}&&
\frac{\partial \rho}{\partial t}+\frac{\partial(\rho v)}{\partial x}=0,\ \
\rho\left(\frac{\partial v}{\partial t}+v\frac{\partial v}{\partial
x}\right)=-\frac{\partial p}{\partial x}, \label{a31} \\&& \frac{\partial
p}{\partial t} +v \frac{\partial p}{\partial x}=-\gamma p\frac{\partial
v}{\partial x}-\Lambda\rho^{1/2} p^{3/2}, \label{a32}
\end{eqnarray}
The family of exact solutions, presented in this Section, are smooth initially
but become singular in a finite time $t_c$. At the singularity, the density
blows up, in contrast to the ordinary ``wave-breaking" singularity of the ideal
gas dynamics ($\Lambda=0$), where only the \textit{gradients} of the
hydrodynamic fields blow up \cite{Landau,Chorin}. An exact solution of a
different type, presented in Section \ref{shocks}, includes a shock already at
$t=0$. That solution also exhibits a density blowup, and the local properties of
the blowup are the same as in the initially smooth solutions.

Let us introduce Lagrangian coordinates. The coordinates $x(m, t)$ of the fluid
particles obey the equation
\begin{equation}
\frac{\partial x(m, t)}{\partial t}=v(x(m, t), t), \label{a51}\end{equation}
where $m$ is a continuous label (a Lagrangian coordinate) of particles. The
defining property of the exact solutions that we are going to present is
independence of the particle accelerations $\partial_t^2 x(m, t)$ of time: the
fluid particle coordinates $x(m, t)$ satisfy the equation
\begin{equation}
x(m, t)=x(m, 0)+v(m, 0) t+\frac{a(m, 0)t^2}{2}, \label{Lagrangacc}\end{equation}
where $v(m, 0)$ and $a(m, 0)$ are the initial velocities and accelerations of
the fluid particles, respectively. As we will see later, the total mass of gas
is finite for our solutions. As a result, the pressure must vanish at the
boundaries, implying existence of a point with a zero pressure gradient (and,
therefore, a zero particle acceleration) in between. A fluid particle that has a
zero acceleration (that is conserved in our solutions) moves with a constant
velocity, and we choose to work in such a frame of reference where this particle
is at rest, so that the conditions $v(x=0, t)=0$ and $\partial_x p(x=0, t)=0$
are obeyed.

It is convenient to choose the Lagrangian coordinate $m$ to be the \textit{mass}
coordinate \cite{zeldovich}:
\begin{equation}
m(x, t)=\int_0^x \rho(x', t)dx'\, \label{masscoordinate}
\end{equation}
that is the mass content between the Eulerian points $0$ and $x$. The inverse
transformation $x(m, t)$ is
\begin{equation}
x(m, t)=\int_0^m \frac{dm'}{\rho(m', t)}\,. \label{space}
\end{equation}
In the Lagrangian coordinates Eqs.~(\ref{a31}) and (\ref{a32}) look
simpler:
\begin{eqnarray}&&
\frac{\partial}{\partial t} \left(\frac{1}{\rho}\right)=
\frac{\partial v}{\partial m},\,\,\, \frac{\partial v}{\partial t}
=-\frac{\partial p}{\partial m},\label{eqs1}\\&& \frac{\partial
p}{\partial t}=-\gamma p \rho\frac{\partial v}{\partial m}-\Lambda
p^{3/2}\rho^{1/2}\,. \label{eqs2}
\end{eqnarray}
Let us calculate the Lagrangian acceleration:
\begin{equation}
\frac{\partial^2 x(m, t)}{\partial t^2}=-\frac{\partial p(m,
t)}{\partial m}, \label{acceleration}
\end{equation}
where we have used Eqs. (\ref{a51}) and (\ref{eqs1}). For the solutions obeying
Eq.~(\ref{Lagrangacc}), the Lagrangian pressure gradient $\partial_m p(m, t)$
should be independent of time. Since the pressure $p(m, t)$ is time-independent
(and zero) at the gas boundaries, it can depend only on $m$. Then
Eqs.~(\ref{eqs1}) and (\ref{eqs2}) yield
\begin{equation}
\frac{\partial^2 p}{\partial m^2}=-\mu^2p,\ \ \ \mbox{where} \ \ \ \mu=
\frac{\Lambda}{\gamma\sqrt{2}}. \label{oscillator}
\end{equation}
In view of the zero acceleration at the origin, $\partial_m p(0, t)=0$,
Eq.~(\ref{oscillator}) yields $p=2A\cos (\mu m)$, where $A$ is constant. The
resulting family of solutions is
\begin{eqnarray}
p(m, t)&=&2A\cos (\mu m), \label{specialp}\\ \rho(m, t)&=&\frac{\rho(m,
0)}{[1-\mu t \sqrt{A\rho(m, 0)\cos (\mu m)}]^2}, \label{specialr}\\ v(m,
t)&=&-2\mu\int_0^m\sqrt{\frac{A\cos (\mu m')}{\rho(m', 0)}}dm'\nonumber
\\&+&2A\mu t\sin(\mu m)\,, \label{specialv}
\end{eqnarray}
where we have used the condition $v(m=0, t)=0$. This family of solutions
describes a compression flow (the velocity gradient is negative everywhere), as
the compressional heating is balanced, in Eq.~(\ref{eqs2}), by the inelastic
cooling. The solutions include an arbitrary non-negative function $\rho(m,0)$:
the initial gas density. The arbitrary constant $A>0$ that, together with
$\rho(m,0)$, determines the (time-dependent) Mach number of the flow, appears
due to the rescaling symmetry of the equations and corresponds to the constant
$\beta$ in Eq.~(\ref{symmetry}). One can check that the constant $\alpha$ in
Eq.~(\ref{symmetry}) corresponds to the freedom of multiplying  $\rho(m,0)$ and
$A$ by $\alpha$.

As the pressure $p$ must be non-negative, and vanish at the (finite or infinite)
boundaries of the freely moving gas, the solutions (\ref{specialp}) can hold
only on a finite interval $(-\pi/2\mu, \pi/2\mu)$ (we assume that the gas region
is single-connected, and the interval includes $m=0$). Therefore, the total mass
of the gas in these solutions is finite and fixed by parameters $\Lambda$ and
$\gamma$: $\int_{-\infty}^{\infty} \rho (x, 0)dx=\pi/\mu=\sqrt{2}
\pi\gamma/\Lambda$.

Once the solutions (\ref{specialp})-(\ref{specialv}) in the Lagrangian
coordinates are known, we can return to the Eulerian coordinates by using, at
any time $t$, Eq. (\ref{space}). Depending on the particular choice of the
initial density, there are two possible types of solutions
(\ref{specialp})-(\ref{specialv}). First, the fixed mass of the gas $\pi/\mu$
can be distributed, at $t=0$, over either an infinite, or a finite $x$-interval.
This is determined by the behavior of $\rho(m, 0)$ near $m=\pm \,\pi/2\mu$. For
example, let $\rho(m, 0)\sim(\pi/2\mu-m)^{1+a}$ near $m=\pi/2\mu$. Then it
follows from Eq. (\ref{space}) that the gas occupies an infinite
(correspondingly, a finite) interval of positive $x$ if $a\geq 0$
(correspondingly, $a<0$). The velocity can be either finite, or infinite at the
gas boundaries. For example, for the same behavior of the initial density
$\rho(m,t=0)\sim(\pi/2\mu-m)^{1+a}$ near $m=\pi/2\mu$ one obtains a finite
(correspondingly, infinite) gas velocity at $m=\pi/2\mu$ for $a<2$
(correspondingly,  $a\geq 2$).

Now let us consider what types of initial conditions evolve according to
Eqs.~(\ref{specialp})-(\ref{specialv}) and discuss the density blowup that is
brought by this evolution.

\subsection{Initial conditions and density blowup for the exact solutions}

A particular member of our family of exact solutions
(\ref{specialp})-(\ref{specialv}) is determined by a specific choice of the
constant $A>0$ and of the initial density $\rho(m, 0)\geq 0$, defined on the
interval $[-\pi/(2\mu), \pi/(2\mu)]$. In the Eulerian coordinates one can
specify an arbitrary initial density profile $\rho(x, 0)$ that has a fixed total
mass $\pi/\mu$:
\begin{equation}
\int_{-\infty}^{\infty}\rho(x, 0)dx=\pi/\mu. \label{masses}
\end{equation}
Once $\rho(x, 0)$ and $A$ are specified, one needs to choose the origin so that
the gas masses to the left and to the right of the origin are the same [and
equal to $\pi/(2\mu)$]. Then the initial gas pressure in the Eulerian
coordinates is
\begin{equation}
p(x, 0)=2A\cos\left(\mu\int_0^x\rho(x', 0)dx'\right).
\label{pressure}
\end{equation}
Now, using Eq.~(\ref{specialv}), we calculate the velocity gradient in the
Eulerian coordinate at $t=0$:
\begin{eqnarray}
\frac{\partial v(x, 0)}{\partial x}&=&\rho(m, 0)\frac{\partial v(m, 0)}{\partial
m}\nonumber \\
&=&-2\mu \sqrt{A\rho(m, 0)\cos (\mu m)}. \label{shear0}\end{eqnarray} which, in
view of the condition $v(x=0,t)=0$, yields
\begin{equation}
v(x, 0)=-\mu\int_0^x \sqrt{2\rho(x', 0)p(x', 0)}dx'
\label{initialvelocity1}\end{equation} with $p(x, 0)$ from Eq.~(\ref{pressure}).
Equations (\ref{pressure}) and (\ref{initialvelocity1}) show that, once $\rho(x,
0)$ is a smooth function of $x$, then the initial pressure and velocity are
smooth functions as well.

Let us now proceed to the properties of the solutions. As we already noted,
these solutions describe a motion of fluid particles with a time-independent
acceleration, see Eq.~(\ref{Lagrangacc}). This time-independent acceleration is
\begin{equation}
a(m, 0)=-\frac{\partial p(m, 0)}{\partial m}=2\mu A\sin (\mu
m).\nonumber
\end{equation}
The evolution described by Eqs. (\ref{specialp})-(\ref{specialv}) brings about a
singularity of this initially smooth flow:  $\rho(m, t)$ blows up in a finite
time, while the rest of the flow variables - the pressure and velocity - remain
finite. The density blow up occurs at the Lagrangian point $m_0$ where $\rho(m,
0)\cos (\mu m)$ reaches its maximum:
\begin{equation} \rho(m_0, 0)\cos (\mu m_0) = \max\left[\rho(m, 0)\cos (\mu m)\right].
\label{defsing}\end{equation} Interestingly, the point $m_0$ corresponds, in
view of Eq.~(\ref{shear0}), to the point of the absolute (negative) minimum of
the velocity gradient  in the Eulerian coordinates, just as in the case of the
free flow  (that is, zero pressure) singularity \cite{Whitham}. The singularity
occurs when the Jacobian of the Lagrangian transformation of the fluid particles
vanishes for the first time: $\partial_m x(m=m_0, t_c)=0$. For the time $t_c$
and the Eulerian coordinate of the singularity $x_0$ we find
\begin{eqnarray}
t_c&=&\frac{1}{\mu \sqrt{A\rho(m_0, 0)\cos (\mu m_0)}}, \nonumber \\
x_c&=&x(m_0, t_c)=\int_0^{m_0} \frac{dm'}{\rho(m', t_c)}\,. \label{critical}
\end{eqnarray}
At the (fixed) Lagrangian point of singularity $m_0$ the density blows up  as
\begin{equation}
\rho(m_0, t)=\frac{\rho(m_0, 0)}{(1-t/t_c)^2}.
\label{densityblowuplagr}\end{equation} In the Eulerian coordinates the blowup
develops, in general, in a moving point:
\begin{equation}
\rho[x(m_0, t), t]=\frac{\rho[x(m_0, 0), 0]}{(1-t/t_c)^2}.
\label{densityblowup}\end{equation} The velocity gradient at the singularity
point $x(m_0,t)$ diverges. The divergence law can be easily found in the
Langrangian coordinates, by using the continuity equation and
Eqs.~(\ref{specialr}) and (\ref{critical}):
\begin{eqnarray}&&
 \left. \frac{\partial v(x,t)}{\partial x} \right|_{x=x(m_0,t)} =
 \left.\rho(m,t) \,
  \frac{\partial v(m,t)}{\partial m}\right|_{m=m_0} \nonumber \\
  &=& \left. -\frac{\partial}{\partial t}\left[\ln \rho (m,t)\right]\right|_{m=m_0}=
  -\frac{2}{t_c-t}\,.\label{velocityderivativeblowup}
\end{eqnarray}
Finally, the (finite) pressure is conserved on the Lagrangian trajectory:
\begin{equation}
p(x=x(m_0, t), t)=2A\cos (\mu m_0). \label{pressureblowup}\end{equation} Note
that the pressure does not have a minimum at the singularity point, so this flow
does not conform to the popular ``pressure instability" scenario
\cite{Goldhirsch}.

It is instructive to consider several particular examples of solutions starting
with the mass distributed over an infinite interval of $x$.

\subsection{Solutions with mass distributed over an infinite interval of $x$}

\begin{figure}
\includegraphics[width=8.5 cm,clip=]{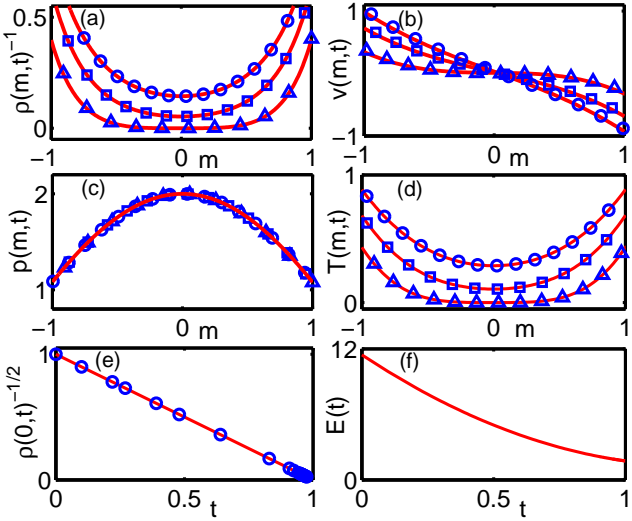}
\caption{(Color online) An example of a single-peak solution in the Lagrangian
coordinates [Eqs.~(\ref{example1b})-(\ref{example1c})], and a numerical solution
for the same parameters. The analytical and numerical solutions are depicted by
solid lines and symbols, respectively. Shown are the inverse density (a),
velocity (b), pressure (c) and temperature (d) at times $t=0.62$ (circles),
$0.77$ (squares) and $0.99$ (triangles) as functions of the Lagrangian
coordinate $m$. Shown in (e) is the inverse square root of the density at the
singularity point $m=0$ (circles). Shown in (f) is the total energy of the gas
versus time as described by Eq.~(\ref{energydecay1}). The density, velocity,
pressure and temperature are rescaled to $\rho_0$, $\sqrt{T_0/2}$, $\rho_0T_0/2$
and $T_0/2$, respectively. The Lagrangian mass coordinate is measured in units
of $M/\pi=1/\mu$. Time is measured in units of $\tau$, so the density blows up
at $t=1$. Details of the numerical solution are given in
Section~\ref{numerics}.}\label{fig1}
\end{figure}

In our first example the initial density profile in the Lagrangian coordinates
is $\rho(m, 0)=\rho_0\cos (\mu m)$. To return to the Eulerian coordinates, we
use Eq.~(\ref{space}) and obtain
\begin{equation}
\sinh \left[\frac{x(m, 0)}{l}\right]=\tan (\mu m)\,, \label{first}\end{equation}
where we have introduced the characteristic inelastic cooling length scale
$l=1/(\mu\rho_0)$ whose meaning will become clear shortly. We use
Eq.~(\ref{first}) to express $\rho(m, 0)=\rho_0\cos (\mu m)$ through $x$. The
rest of initial conditions follow from Eq.~(\ref{specialp}) and
Eq.~(\ref{specialv}) at $t=0$. Note that the initial gas temperature $T(m,
0)=p(m, 0)/\rho(m, 0)=2A/\rho_0\equiv T_0$ is uniform in space. The initial
conditions are
\begin{eqnarray}&&
\rho(x, 0)=\frac{\rho_0}{\cosh (x/l)},\ \ T(x,
0)=T_0,\label{initial0}
\\&& v(x, 0)=-\sqrt{2T_0}\arcsin
\left[\tanh \left(\frac{x}{l}\right)\right]. \label{initial}\end{eqnarray} The
initial velocity profile describes an inflow of gas from plus and minus infinity
with a finite velocity there: $\lim_{x\to \pm \infty}v(x, 0)=\mp\,
\pi\sqrt{T_0/2}$. Now let us introduce the characteristic inelastic cooling time
\begin{equation}
\tau= \frac{l\sqrt{2}}{\sqrt{T_0}} =\frac{\sqrt{2}}{\mu\rho_0\sqrt{T_0}}=\frac{2
\gamma}{\rho_0 \Lambda \sqrt{T_0}}\,,
\end{equation}
see Eq.~(\ref{a322}). In its turn, $l$ is the characteristic length scale the
particles pass during the time $\tau$ while moving with thermal velocity.
According to Eqs.~(\ref{specialp})-(\ref{specialv}), the hydrodynamic fields in
the Lagrangian coordinates evolve in time in this example as
\begin{eqnarray}
p(m, t)&=&\rho_0T_0\cos (\mu m),\label{example1b}\\
\rho(m, t)&=&\frac{\rho_0\cos (\mu m)}{[1-(t/\tau)\cos (\mu m)]^2},\label{example1a} \\
v(m, t)&=&-\sqrt{2T_0}\left[\mu m-(t/\tau)\sin (\mu m)\right], \label{example1c}
\end{eqnarray}
as depicted in Fig.~\ref{fig1}.

\begin{figure}
\includegraphics[width=8.5 cm,clip=]{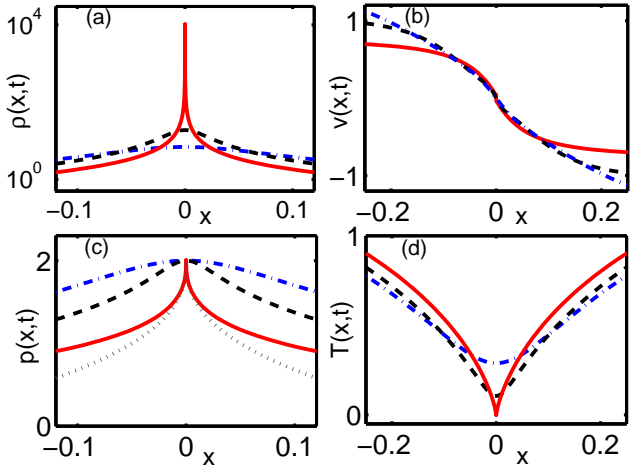}
\caption{(Color online) The exact solution from Fig. \ref{fig1} in the Eulerian
coordinates. Shown are: the density in the logarithmic scale (a), the velocity
(b), the pressure (c) and the temperature (d) at rescaled times $t=0.62$ (the
dashed-dotted line), $0.77$ (the dashed line), $0.99$ (the solid line) and the
blowup time $1$ (the dotted line in c) as functions of $x$. The $x$-coordinate
is measured in units of $l$. The rest of units are the same as in
Fig.~\ref{fig1}.}\label{fig2}
\end{figure}

Using Eqs.~(\ref{space}) and (\ref{example1a}) we find the law of motion
(\ref{Lagrangacc}) of Lagrangian particles,
\begin{eqnarray}
\frac{x(m, t)}{l}&=&\frac{1}{2}\ln\left(\frac{1+\sin (\mu m)}{1-\sin (\mu
m)}\right)\nonumber \\
&-&2\left(\frac{t}{\tau}\right)\mu m+\left(\frac{t}{\tau}\right)^2\sin(\mu m)
\nonumber
\end{eqnarray}
which, combined with Eqs.~(\ref{example1a})-(\ref{example1c}), yields a
parametric representation of the solution in the Eulerian coordinates, see
Fig.~\ref{fig2}. The density singularity occurs at $x=0$ at time $t=\tau$:
\begin{eqnarray}&&
\rho(0, t)=\frac{\rho_0}{(1-t/\tau)^2},\ \ T(0, t)=T_0(1-t/\tau)^2,\nonumber
\\&& p(0, t)=\rho_0T_0=\mbox{const},\ \ \frac{\partial v}{\partial
x}(0,t)=-\frac{2}{\tau-t},\label{singularity}
\end{eqnarray}
while $v(x=0,t)=0$. Notice that the pressure has a local \textit{maximum} at the
density blowup point $x=m=0$.

\begin{figure}
\includegraphics[width=8.5 cm,clip=]{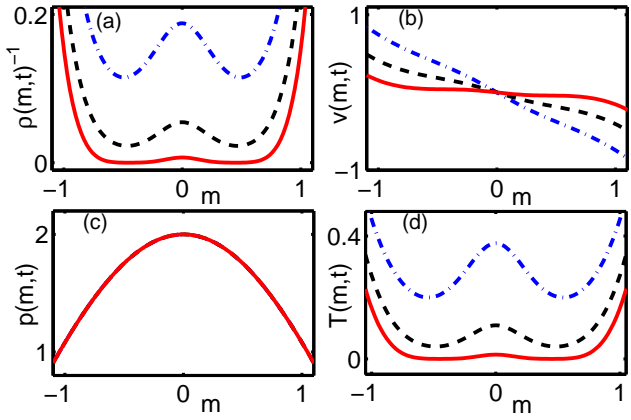}
\caption{(Color online) An example of a two-peak solution in the Lagrangian
coordinates. Shown are the exact solutions~(\ref{specialp})-(\ref{specialv})
with the initial density from Eq.~(\ref{Lagrangiandensity}): the inverse density
(a), velocity (b), pressure (c) and temperature (d) at the rescaled times
$t=1.1$ (the dashed-dotted line), $1.3$ (the dashed line) and $1.45$ (the solid
line) versus the Lagrangian coordinate $m$. The density blows up at the
Lagrangian points $\;\pm \arccos \left(\sqrt{2/3}\,\, \coth 1.5\right)=\pm
0.44628\dots$ at the rescaled time $t_c= 3^{3/4}\,2^{-1/2}\tanh 1.5
=1.45896\dots$. The units are the same as in Fig.~\ref{fig1}.}\label{fig3}
\end{figure}

\begin{figure}
\includegraphics[width=8.5 cm,clip=]{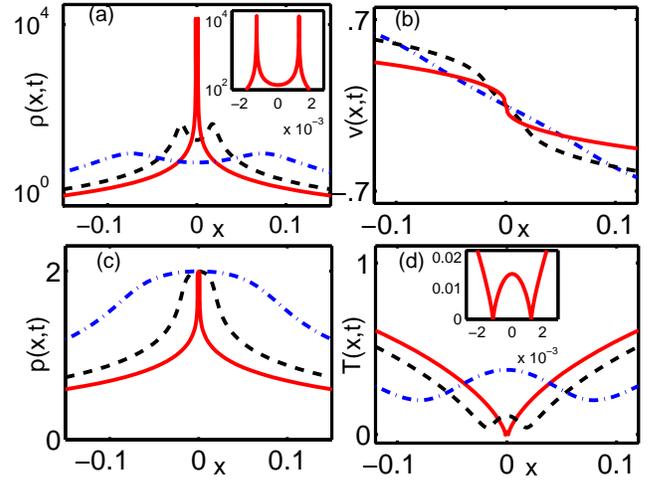}
\caption{(Color online) The exact solution from Fig. \ref{fig3} in the Eulerian
coordinates. Shown are the density in the logarithmic scale (a), velocity (b),
pressure (c) and temperature (d) at rescaled times $t=1.1$ (the dashed-dotted
line), $1.3$ (the dashed line) and $1.45$ (the solid line) as functions of $x$.
The insets in (a) and (d) zoom on the density and temperature profiles,
respectively, at $t=1.45$ and clearly show the presence of two symmetric peaks
at $x \neq 0$ developing density blowups simultaneously. The units are the same
as in Figs.~\ref{fig1} and~\ref{fig2}.}\label{fig4}
\end{figure}

The above solution can be immediately generalized. Indeed, it is a particular
case of a one-parameter family of solutions generated by the initial density
profile
\begin{equation}
\rho(x,
0)=\frac{\rho_0}{2}\left[\cosh^{-1}\left(\frac{x}{l}+a\right)+\cosh^{-1}\left(\frac{x}{l}-a\right)\right],
\label{indens}\end{equation} where $a>0$ is an arbitrary parameter. For
$a<a_{cr}= \mbox{arcsinh}(1)=0.88137\dots$ there is a single density peak at
$x=0$, while at $a>a_{cr}$ there are two symmetric density peaks at $x=\pm\, l\,
\mbox{arccosh}(\sinh a)$. The density profile (\ref{indens}) obeys
Eq.~(\ref{masses}): the total mass of the gas remains equal to $\pi/\mu$.
Equations~(\ref{space}) and (\ref{indens}) yield
\begin{equation}
\sinh \left[\frac{x(m, 0)}{l}\right]=\cosh a \tan (\mu m).
\label{connection}\end{equation} By setting $p(x, 0)=\rho_0T_0\cos [\mu m(x,
0)]$ we obtain the initial temperature
\begin{equation}
T(x, 0)=T_0\sqrt{1+\frac{\sinh ^2 a}{\cosh^2(x/l)}},
\end{equation}
while $v(x, 0)$ can be found from Eq.~(\ref{initialvelocity1}). Now we calculate
the initial conditions in the Lagrangian coordinates. Using Eqs.~(\ref{indens})
and (\ref{connection}) we obtain
\begin{equation}
 \rho(m, 0)=\rho_0\cos(\mu m)\sqrt{1-\tanh^2 a\cos^2 (\mu m)}.
\label{Lagrangiandensity}\end{equation} Then Eq. (\ref{specialv}) yields
\begin{equation}
v(m, 0)=-\sqrt{2T_0}\int_0^{\mu m} \!\frac{dm'}{(1-\tanh^2 a\cos^2 m')^{1/4}}.
\label{initv}
\end{equation}
Though this integral can be expressed via the Appell hypergeometric function of
two variables, the integral form is more visual. The time history of the
hydrodynamic fields in the Lagrangian coordinates is shown in Fig.~\ref{fig3}.
To go over to the Eulerian coordinates, we calculate the law of motion
(\ref{Lagrangacc}) of the Lagrangian particles:
\begin{eqnarray}
\frac{x(m, t)}{l}&=&\ln\left[\cosh a\tan(\mu m)+\sqrt{1+\cosh^2a\tan^2(\mu
m)}\right] \nonumber\\
&-&2\left(\frac{t}{\tau}\right)\int_0^{\mu m} \frac{dm'}{(1-\tanh^2 a\cos^2
m')^{1/4}} \nonumber \\
&+&\left(\frac{t}{\tau}\right)^2\sin (\mu m),\nonumber
\end{eqnarray}
and use it together with the Lagrangian solutions, see Fig.~\ref{fig4}.

The development of singularity in this case depends on the parameter $a$. For
$a<a_{cr}$ the density and the pressure peaks remain at $x=0$ at all times until
the singularity, while the singularity is of the same type as that observed for
$a=0$:
\begin{eqnarray}&&
\rho(0, t)=\frac{\rho_0}{(\sqrt{\cosh a}-t/\tau)^2},\nonumber \\
&&T(0, t)=T_0(\sqrt{\cosh a}-t/\tau)^2 ,\nonumber
\\&& p(0, t)=\rho_0T_0,\ \ \frac{\partial v}{\partial x}=-\frac{2}{\tau\sqrt{\cosh
a}-t}. \label{singularity1}
\end{eqnarray}
The density blowup time is $t_c=\tau\sqrt{\cosh a}$. Now let us consider the
case of $a>a_{cr}$ with two symmetric off-center density peaks at $t=0$.
Interestingly, at $a_{cr}<a<{\tilde a}_{cr}$, where ${\tilde
a}_{cr}=\mbox{arcsinh}(\sqrt{2})=1.14621\dots$, the singularity still develops
at $x=0$. Indeed, it is the maximum of the function $\rho(m, 0)\cos (\mu m)$,
rather than of $\rho(m,0)$ that determines, in view of Eq.~(\ref{defsing}), the
singularity point. For $a_{cr}<a<{\tilde a}_{cr}$ this maximum is still at
$m=m_0=0$. As time progresses, the two symmetric density peaks move toward the
origin, reaching it precisely at the time of singularity. The pressure still has
a maximum at $x=0$, while in view of Eqs.~(\ref{singularity1}) the time of
singularity is still $t_c=\tau \sqrt{\cosh a}$. Instead of looking for the
maxima of $\rho(m)$, it is convenient to look for the minima of the inverse
density which, by virtue of Eq.~(\ref{specialr}), can be written as
\begin{eqnarray}&&
\frac{\rho_0}{\rho(m, t)}=\frac{1}{\cos (\mu m)\sqrt{1-\tanh^2 a\cos^2(\mu
m)}}-2\left(\frac{t}{\tau}\right)\nonumber\\&&\times\frac{1}{\left[1-\tanh^2
a\cos^2 (\mu m)\right]^{1/4}}+\left(\frac{t}{\tau}\right)^2\cos (\mu m).
\end{eqnarray}
Differentiating this  function with respect to $m$ one can verify that at
$t=\tau\sqrt{\cosh a}$ the minimum is indeed at $m=0$. Finally, at
$a>{\tilde{a}_{cr}}$ the density blows up symmetrically at two Eulerian points
$x=\pm\, x(m_0, t_c)\neq 0$, where
$$
m_0=(1/\mu)\,\arccos\left(\sqrt{2/3}\,\, \coth a\right) \neq 0.
$$
Using the first of Eq.~(\ref{critical}) we find $t_c/\tau = 3^{3/4}\,
2^{-1/2}\tanh a$. In this case the pressure gradient is non-zero in the
singularity point, so the pressure is neither maximum, nor minimum.

\subsection{Solutions with mass distributed over a finite interval of $x$}

If $\rho(m, 0)$ does not vanish at $|m|=\pi/(2\mu)$, or vanishes slower than
linearly there, then the integral in Eq.~(\ref{space}) converges, and the total
gas mass $\pi/\mu$ is distributed over a finite interval of $x$. It follows from
Eq.~(\ref{specialv}) that the velocity (which vanishes at $x=m=0$ and has a
negative gradient everywhere) takes finite values at the ends of the interval
over which the mass is distributed. As a result, the $x$-interval, occupied by
the gas, shrinks in the course of evolution. Assuming for simplicity that
$\rho(m, 0)$ is an even function of $m$, we find from Eqs.~(\ref{space}) and
(\ref{specialr}) that the interval $\left[-L(t), L(t)\right]$, occupied by the
gas, shrinks with time as
\begin{equation}
L(t)=L(0)-2\mu t\int_0^{\pi/2\mu}\sqrt{\frac{A\cos (\mu m')}{\rho(m',
0)}}\,dm'+\mu A t^2 \nonumber
\end{equation}
and reaches its minimum at $t=t_c$. This minimum is always positive except in
the degenerate case of $\rho(m,0)=\rho_0/\cos(\mu m)$, when the whole gas
collapses into the point $x=0$ at the time of singularity. It turns out that, in
this degenerate case, the solution is self-similar in the Eulerian coordinates
$x$ and $t$ and separable in the Lagrangian coordinates $m$ and $t$. It can be
shown that all self-similar solutions with a finite energy have an infinite
density at some locations already at $t=0$. Such initial conditions do not
correspond to a dilute gas, so they will not be considered here.

A generic example of the solution on a finite Eulerian interval is provided by
the uniform initial density $\rho(m, 0)=\rho_0$. In the Eulerian coordinates
this choice corresponds to the constant initial density, $\rho(x, 0)=\rho_0$, at
the interval $[-\pi l/2, \pi l/2]$. The solutions
(\ref{specialp})-(\ref{specialv}) become
\begin{eqnarray}
p(m, t)&=&\rho_0T_0\cos (\mu m),\label{example2b}\\
\rho(m, t)&=&\frac{\rho_0}{[1-(t/\tau)\sqrt{\cos (\mu m)}]^2},\label{example2a} \\
v(m, t)&=&-\sqrt{2T_0}\left[2 \mathrm{E}\left.\left(\frac{\mu m}{2}\,\right|
\,2\right) -\frac{t}{\tau}\sin (\mu m)\right], \label{example2c}
\end{eqnarray}
where $\mathrm{E}\left(\dots |\dots\right)$ is the elliptic integral of the
second kind. The relation for $x(m,t)$ is the following:
\begin{eqnarray}
\frac{x(m, t)}{l}&=&\mu m - 4\left(\frac{t}{\tau}\right)
\mathrm{E}\left.\left(\frac{\mu m}{2}\,\right|
\,2\right)+\left(\frac{t}{\tau}\right)^2 \sin (\mu m).\nonumber
\end{eqnarray}
Here the singularity occurs at $x=0$ at time $t_c=\tau$ whereas $L(t_c)>0$. The
local structure of singularity is the same as in the case of an infinite
interval, see subsection~\ref{local}.

\subsection{Energy decay for the exact solutions}
\label{energy}

A useful \textit{global} characteristics of the clustering process is provided
by the evolution of the total kinetic energy of all particles versus time,
$E(t)>0$. This quantity is convenient to follow in experiment and in MD
simulations. In the language of hydrodynamics it is
\begin{equation}
E(t)=\int_{-\infty}^{\infty}  \left(\frac{\rho T}{\gamma-1}+\frac{\rho v^2}{2}
\right)dx\,,
\end{equation}
where the first term under the integral is the thermal energy density, and the
second term is the macroscopic kinetic energy density. Let us compute $E(t)$ for
the exact solutions Eqs.~(\ref{specialp})-(\ref{specialv}). First, consider the
conditions under which the initial total energy $E(0)$ is finite.  For the
initial thermal energy we have
\begin{eqnarray}
  E_{th}(0) &=& \int_{-\infty}^{\infty}\frac{\rho(x,
0) T(x, 0)dx}{\gamma-1}  \\ \nonumber
   &=& 2A
\int_{-\pi/2\mu}^{\pi/2\mu} \frac{\cos(\mu m)dm}{(\gamma-1)\rho(m, 0)}\,.
\end{eqnarray}
Whether this quantity is finite or not depends on the behavior of $\rho(m, 0)$
near $m=\pm \pi/2\mu$. For example, assuming as before that $\rho(m,
0)\sim(\pi/2\mu-m)^{1+a}$ near $m=\pi/2\mu$, we find that $E_{th}(0)$ is finite
at $a<1$ and infinite otherwise. A simple example of the initial condition with
an infinite energy is $\rho(m, 0)=\rho_0\cos^2(\mu m)$ corresponding to the
Lorentzian initial density profile $\rho(x, 0)=\rho_0/[1+(x/l)^2]$  in the
Eulerian coordinates. Here
$$\rho(x, 0)T(x, 0)=\rho_0T_0/\sqrt{1+(x/l)^2}\,,\;\;\;-\infty<x<\infty\,,$$
which is not integrable. Now, since at $a<2$ the gas velocity is finite at
$m=\pi/2\mu$ (see subsection \ref{Lagrangian}), then at $a<1$ the initial
macroscopic kinetic energy
\begin{equation}
E_{kin}(0) = \int_{-\infty}^{\infty}\frac{\rho v^2}{2} dx =
\int_{-\pi/2\mu}^{\pi/2\mu} \frac{v^2(m, 0)}{2} dm
\end{equation}
is also finite. Therefore, we assume $a<1$ so that $E(0)<\infty$. Using the
divergence form of the energy equation,
\begin{eqnarray}
\frac{\partial}{\partial t}\left(\frac{\rho T}{\gamma-1}+\frac{\rho
v^2}{2}\right)&+&\frac{\partial }{\partial x}\left(\frac{\gamma\rho v
T}{\gamma-1}+\frac{\rho v^3}{2}\right) \nonumber \\  &=&-\frac{\Lambda \rho^2
T^{3/2}}{\gamma-1}\,,
\end{eqnarray}
we can express the decay rate of the total energy as
\begin{eqnarray}
\frac{dE}{dt}&=&-\frac{\Lambda}{\gamma-1}\int_{-\infty}^{\infty}
\rho^2T^{3/2}dx \nonumber \\
&=&-\frac{\Lambda}{\gamma-1}\int_{-\pi/2\mu}^{\pi/2\mu}\frac{p^{3/2}
dm}{\rho^{1/2}}\,.
\end{eqnarray}
Using Eqs. (\ref{specialp}) and (\ref{specialr}), we obtain
\begin{equation}
\frac{dE}{dt}=-\frac{\Lambda(2A)^{3/2}}{\gamma-1}
\left[\int_{-\pi/2\mu}^{\pi/2\mu} \frac{dm \cos^{3/2}(\mu m)}{\sqrt{\rho(m,
0)}}-\frac{\pi \sqrt{A} t}{2} \right]. \nonumber
\end{equation}
It is easy to see that the integral in this equation converges for the assumed
behavior of the initial density. Therefore, the energy decays quadratically in
time:
\begin{eqnarray}
E(t)&=&E(0)-\frac{\Lambda(2A)^{3/2}t}{\gamma-1} \int_{-\pi/2\mu}^{\pi/2\mu}
\frac{dm \cos^{3/2}(\mu m)}{\sqrt{\rho(m, 0)}}\nonumber \\
&+&\frac{\pi \Lambda A^2t^2}{\sqrt{2}(\gamma-1)}. \label{energydecay}
\end{eqnarray}
One can check that the decay of the thermal and macroscopic kinetic energies,
separately, is also quadratic in time. One also observes that, at the time of
the density blowup, $t=t_c$, the energy constitutes a finite and non-zero
fraction of the initial energy, so the time $t_c$ is not special for the
function $E(t)$.

As a simple example, consider the initial density $\rho(m, 0)=\rho_0\cos(\mu m)$
corresponding to Eqs.~(\ref{initial0}) and (\ref{initial}). Here the integration
in Eq.~(\ref{energydecay}) is elementary, and we obtain
\begin{equation}
\frac{E(t)}{\rho_0T_0l}=\frac{\pi}{\gamma-1}+\frac{\pi^3}{12}-\frac{4\gamma}{\gamma-1}\left(
\frac{t}{\tau}\right)+\frac{\pi\gamma}{2(\gamma-1)}\left(
\frac{t}{\tau}\right)^2\,,\label{energydecay1}
\end{equation}
where $\rho_0T_0 l$ is the characteristic energy scale. This $E(t)$ dependence
is shown in Fig.~\ref{fig1}. We now proceed to the analysis of the
\textit{local} structure of the flow near a singularity.

\subsection{Local structure of the exact solutions near the singularity} \label{local}

To analyze the local structure of the developing singularity we consider the
Taylor expansion of the inverse density $u(m, t)\equiv 1/\rho(m, t)$ in a close
vicinity of $m=m_0$ at times close to $t_c$. To calculate the $m$-derivatives it
is convenient to write $u(m, t)=u(m, 0)[1-t\phi(m)]^2$, where $\phi(m)= \mu
\sqrt{A\rho(m, 0)\cos (\mu m)}$ satisfies the conditions $\phi(m_0)=1/t_c$ and
$\phi'(m_0)=0$, see Eqs.~(\ref{defsing}) and (\ref{critical}). After some
algebra we find
\begin{eqnarray}
\frac{u'(m_0, t)}{u'(m_0, 0)}&=&\Delta^2,\nonumber \\
\frac{u''(m_0, t)}{u(m_0, 0)}&=&-2t_c\phi''\Delta+{\cal O}(\Delta^2), \nonumber \\
u'''(m_0, t)&=&-2 t_c\Delta\left[3u'(m_0, 0)\phi''+u(m_0,
0)\phi'''\right]\nonumber\\&&+{\cal O}(\Delta^2), \nonumber \\
\frac{u^{(4)}(m_0, t)}{u(m_0, 0)}&=&6 t_c^2\phi''^2+{\cal O}(\Delta),
\label{taylorcoeff}
\end{eqnarray}
where all the derivatives of $\phi$ are evaluated at $m=m_0$, and
$\Delta\equiv 1-t/t_c$. The first non-vanishing $m$-derivative at $t=t_c$ is,
therefore, of the fourth order, and we obtain
\begin{equation}
\frac{u(m, t_c)}{u(m_0, 0)}\simeq b^2\mu^4(m-m_0)^4,\ \ \mu|m-m_0|\ll 1,
\label{quartic}\end{equation} where we have introduced a positive dimensionless
constant $b= -t_c\phi''(m_0)/2\mu^2$ [recall that $\phi(m)$ has a maximum at
$m=m_0$, so that $\phi''(m_0)<0$]. Using the expression for $t_c$ from Eq.
(\ref{critical}) and the definition of $\phi$ one finds that $b$ is independent
of $A$ and can be written as
$$b=-\frac{1}{4\mu^2}\left.\frac{d^2}{dm^2} {\ln\left[\rho(m, 0)\cos(\mu
m)\right]}\right|_{m=m_0},$$ where we have used $\left[\rho(m, 0)\cos (\mu
m)\right]'(m=m_0)=0$. The latter relation implies
$(\rho'/\rho)(m=m_0)=\mu\tan(\mu m_0)$, so we obtain
\begin{equation}
b=\frac{1}{4}\left[1+2\tan^2 m_0-\frac{\rho''(m_0, 0)}{\mu^2\rho(m_0, 0)}\right]
={\cal O}(1)\,,
\end{equation}
where the latter estimate assumes that the initial density $\rho(m, 0)$ varies
over a scale of order $1/\mu$. Equation~(\ref{quartic}) shows that, at the
singularity, $u=1/\rho$ vanishes faster than quadratically in $m$, as expected
in general when a singularity is analytic. Going back to the Eulerian
coordinates, $x(m, t)-x(m_0, t)=\int_{m_0}^m u(m', t)dm'$, we rewrite
Eq.~(\ref{quartic}) as
\begin{equation}
\frac{u(x, t_c)}{u(x_0, 0)}\simeq \left[\frac{5
\sqrt{b}(x-x_c)}{l}\right]^{4/5},\ \left|\frac{x-x_c}{l}\right|^{1/5}\ll 1,
\label{usingular}
\end{equation}
where $x_0\equiv x(m_0, 0)$, the spatial coordinate of the singularity $x_c$ is
determined by Eq. (\ref{critical}), and $l=1/[\mu\rho(x_0, 0)]$ is the inelastic
cooling length scale. The validity condition in $x$ in Eq.~(\ref{usingular})
corresponds to the validity condition in $m$ in Eq.~(\ref{quartic}). Thus, at
$t=t_c$ the density profile is singular in a vicinity of $x=x_c$, and exhibits a
power law with exponent $4/5$:
\begin{equation}
\frac{\rho(x, t_c)}{\rho(x_0, 0)}\simeq
\left(\frac{l}{5\sqrt{b}\,|x-x_c|}\right)^{4/5}. \label{power}\end{equation}
This power-law singularity is integrable (that is, has a finite mass) and
symmetric with respect to $x_c$. For comparison, the density singularity of a
free flow, see \textit{e.g.} \cite{Whitham}, exhibits the exponent $2/3$ instead
of $4/5$ .

As the velocity itself is finite at singularity, the velocity gradient is of
interest. We obtain
\begin{equation}
\left.\frac{\partial v}{\partial x}\right|_{x=x(m, t)}=\rho(m, t)\frac{\partial
v}{\partial m}=-\frac{2\phi(m)}{1-t\phi(m)}\,,
\end{equation}
where the last equality follows from the continuity equation and definition of
$\phi(m)$. As a result,
\begin{equation}
-\frac{2}{\partial_xv}=\frac{1}{\phi(m)}-t\simeq
t_c-t+bt_c\mu^2(m-m_0)^2\,,\label{criticalvelocity}
\end{equation}
up to higher order terms in $t_c-t$ and $m-m_0$. Using
Eqs.~(\ref{criticalvelocity}) and (\ref{quartic}), we find that
$$
-\frac{2}{t_c\partial_x v(x,t_c)} =\sqrt{\frac{u(m, t_c)}{u(m_0, 0)}}\,.
$$
This relation, combined with Eq.~(\ref{usingular}), yields
\begin{equation}
\frac{\partial v(x,t_c)}{\partial x} = - \frac{2}{t_c}\left[\frac{l}
{5\sqrt{b}\,(x-x_c)}\right]^{2/5}, \label{velocitypower}\end{equation} Note that
while the exponent $2/5$ of this power law is different from the exponent $4/5$
of the power law for the density, the two power laws have the same region of
validity in $x$, see Eq.~(\ref{usingular}). Though the velocity gradient
diverges at the singularity, the velocity itself is continuous there and has a
cusp  $\sim |x_c-x|^{3/5}$. This is in contrast with the ``wave breaking"
singularity of ordinary gas dynamics, where the velocity becomes discontinuous
at the point where the velocity gradient blows up.

It is clear from the above that the local profiles of the density and velocity
at $t=t_c$ do not depend on the details of the initial density $\rho(m, 0)$.
This is not so for the pressure for which two types of spatial behavior are
possible. For the special case where $\rho(m, 0)\cos (\mu m)$ is maximum at
$m=0$, and so $m_0=0$, the density blows up at the origin $x=0$. As a result,
the pressure versus $x$ has a local maximum at $x=0$, and the value of the
maximum stays constant in time: $p(x=0, t)=2A$. The pressure profile in a
vicinity of $x=0$ can be obtained from a Taylor expansion of
Eq.~(\ref{specialp}): $p(m, t)/A\simeq 2-\mu^2m^2+{\cal O}(\mu^4m^4)$. At
$t=t_c$ this leads to
\begin{equation}
\frac{p(x, t_c)}{p(0, t_c)}\simeq 1-\frac{1}{2}\left(\frac{5 x}{l
b^2}\right)^{2/5}\,,\;\;\;\;\;\;\left(\frac{x}{l }\right)^{2/5}\ll
1\,.\label{pressurezero}
\end{equation}
In the generic case, where the maximum of $\rho(m, 0)\cos (\mu m)$ is not at
zero, so that $m_0\neq 0$, the singularity develops at a point which is not
special for the pressure. Here the Taylor expansion of Eq.~(\ref{specialp})
yields
\begin{eqnarray}
  p(m)/A &=& 2\cos(\mu m_0)-2\mu(m-m_0)\sin(\mu m_0) \nonumber \\
  &+& {\cal O}[\mu^2(m-m_0)^2]\,.
\end{eqnarray}
In this case we find
\begin{equation}
\frac{p(x_c, t_c)}{p(x_c, t_c)}\simeq 1-\tan(\mu
m_0)\left[\frac{5(x-x_c)}{lb^2}\right]^{1/5} ,
\label{pressurenzero}\end{equation} that holds at
$(|x-x_c|/l)^{1/5}\ll 1$. Note that the local $x$-dependence of the
pressure at $t=t_c$ is very different from that of the density, or
velocity gradient. First, the pressure remains finite at $t=t_c$.
Second, even though the pressure gradient diverges at $x=x_c$, the
divergence stems from a small correction term to a constant
pressure, see Eq.~(\ref{pressurezero}) and~(\ref{pressurenzero}). As
we discuss later, this difference in behavior is crucial for
understanding the physical nature of the singularity.

Now let us investigate the local structure of the flow immediately before the
singularity: at $t_c-t\ll t_c$. The leading terms of the double Taylor expansion
of the inverse density $u(m, t)=u(m, 0)[1-t\phi(m)]^2$ in a vicinity of $m=m_0$
and $t=t_c$ (that is, at $\mu|m-m_0|\ll 1$ and $\Delta\ll 1$) are the following:
\begin{equation}
\frac{u(m, t)}{u(m_0, 0)}\simeq \Delta^2+2b\Delta \mu^2
(m-m_0)^2+b^2\mu^4(m-m_0)^4\,;\label{Taylor}
\end{equation}
we recall that $\Delta=1-t/t_c$. Equation~(\ref{Taylor}) can be written in a
self-similar form:
\begin{equation}
\frac{u(m, t)}{u(m_0, 0)}=\left(1-\frac{t}{t_c}\right)^2
U\left[\frac{\mu(m-m_0)}{\sqrt{1-t/t_c}} \right]\,,
\end{equation}
where $U(y)=(1+b y^2)^2$. Now, $x(m, t)-x_c(t)=\int_{m_0}^m u(m', t)dm'$ can be
written as
$$
\frac{x(m, t)-x_c(t)}{l}=\left(1-\frac{t}{t_c}\right)^{5/2} \Xi
\left[\frac{\mu(m-m_0)}{\sqrt{1-t/t_c}} \right]\,,
$$
where $\Xi(y)=\int_0^y U(y')dy'$. Evaluating this integral, we arrive at the
self-similar form in the Eulerian coordinates:
\begin{equation}
\frac{u(x, t)}{u(x_0, 0)}=\left(1-\frac{t}{t_c}\right)^{2}\left\{
1+bF^2\left[\frac{x-x_c(t)} {l(1-t/t_c)^{5/2}}
\right]\right\}^2,\label{physpace}
\end{equation} where the function $F(z)$ is defined implicitly by
the fifth order polynomial equation
\begin{equation}
\frac{b^2F^5(z)}{5}+\frac{2bF^3(z)}{3}+F(z)=z
\end{equation}
that has a unique real solution. Equation~(\ref{physpace}) holds at $x$ that
satisfy the strong inequality
\begin{equation}\label{criter}
    \left(1-\frac{t}{t_c}\right)^{1/2}
\left|F\left[\frac{x-x_c(t)}{l(1-t/t_c)^{5/2}}\right] \right| \ll 1\,,
\end{equation}
corresponding to the condition $\mu|m-m_0|\ll 1$ in Eq. (\ref{Taylor}). The
asymptotes of $F(z)$ are
\begin{equation}
F (z)\simeq\left\{\begin{array}{ll}
z\,,\;\;\;\;\;\;\;\;\;\;\;\;\;\;\;\;\;\; z^2\ll 1\,, \nonumber\\
\\
(5z/b^2)^{1/5}\,,\;\;\;\;z^{2/5}\gg 1\,.\end{array}\right.
\end{equation}
The applicability condition (\ref{criter}) is determined by the $|z|\gg1$
asymptote of $F(z)$ and simplifies to
$$
\left|\frac{x-x_c(t)}{l}\right|^{1/5}\ll 1\,.
$$
The same condition guarantees the validity of the power-law density profile
(\ref{power}) at $t=t_c$.

Therefore, near the singularity the density has the following self-similar form:
\begin{equation}
\frac{\rho(x, t)}{\rho(x_0, 0)}=\frac{1}{(1-t/t_c)^2}R\left[\frac{x-x_c(t)}
{l(1-t/t_c)^{5/2}} \right],\nonumber
\end{equation}
where $R(z)=1/[1+bF^2(z)]^2$. In the region corresponding to $z\ll 1$ one has
$R(z)\simeq 1$. That is, $\rho(x, t)$ develops a plateau in a narrow region near
$x_c(t)$ that we will call the \textit{inner} region. The inner region shrinks
as $(t_c-t)^{5/2}$ as $t$ approaches $t_c$:
\begin{equation}
\rho(x, t)\simeq \frac{\rho(x_0, 0)}{(1-t/t_c)^2}\;\;\;\mbox{at} \;\;\;\;
\left[\frac{\left|x-x_c(t)\right|}{l(1-t/t_c)^{5/2}}\right]^2\ll 1.\nonumber
\end{equation}
At intermediate distances, or in the \textit{outer} region,
$$
\left[\frac{\left|x-x_c(t)\right|}{l(1-t/t_c)^{5/2}}\right]^{2/5}\gg 1,\ \
\left|\frac{x-x_c(t)}{l}\right|^{1/5}\ll 1 \,,
$$
the time-independent power law (\ref{power}) builds up. At $t=t_c$
the power law rules in the whole region $(|x-x_c(t)|/l)^{1/5}\ll 1$.

The development of the singular profile of the velocity gradient can be inferred
by going back to Eq.~(\ref{criticalvelocity}):
\begin{equation}
\frac{\partial v}{\partial x}=-\frac{2}{(t_c-t)[1+bF^2(z)]},\;\;\;\;
z=\frac{[x-x_c(t)]} {l(1-t/t_c)^{5/2}}.\nonumber
\end{equation}
Thus $\partial_x v$ develops a plateau $\partial_x v=-2/(t_c-t)$ [cf. Eq.
(\ref{velocityderivativeblowup})] in the inner region, while the power law
described by Eq. (\ref{velocitypower}) sets in the outer region in the same way
as the density power law.

The development of the pressure profile as $t$ approaches $t_c$ is different in
the case of $m_0=0$ and $m_0\neq 0$. In the former case we have
\begin{equation}
\frac{p(x, t)}{p(0, t)}\simeq 1-\frac{1-t/t_c}{2} F^2\left[\frac{x}
{l(1-t/t_c)^{5/2}} \right],\nonumber
\end{equation}
that holds at $(|x|/l)^{2/5}\ll 1$. In the inner region,
$\left[|x|/l(1-t/t_c)^{5/2}\right]^2\ll 1$,
\begin{eqnarray}&&
\frac{p(x, t)}{p(0, t)}\simeq 1-\frac{1}{(1-t/t_c)^4}\frac{x^2}{2l^2},
\label{seconder}\end{eqnarray} while in the outer region the time-independent
profile (\ref{pressurezero}) sets it.

In the generic case of $m_0\neq 0$ we obtain, for
$\left[|x-x_c(t)|/l\right]^{1/5}\ll 1$,
\begin{eqnarray}
\frac{p(x, t)}{p[x_c(t), t]}&\simeq& 1 \nonumber \\
&-&\tan (\mu m_0)\sqrt{1-t/t_c}\, F\left[\frac{x-x_c(t)} {l(1-t/t_c)^{5/2}}
\right]\,. \nonumber
\end{eqnarray}
In the inner region this yields
\begin{equation}
\frac{p(x, t)}{p[x_c(t), t]}\simeq1-\tan(\mu m_0) \frac{x-x_c(t)}
{l(1-t/t_c)^{2}}\,, \nonumber
\end{equation}
whereas the time-independent profile~(\ref{pressurenzero}) sets in the outer
region. We observe that in the inner region, $\left[|x-x_c(t)|/l\right]^{1/5}\ll
1$, the pressure is approximately constant. This suggests that the sound waves
are the fastest physical process near the singularity. Before we consider the
hierarchy of time scales in more detail, let us reiterate that the finite-time
singularity described here is quite different from the free-flow singularity
\cite{Whitham} where the density blows up as $(t_c-t)^{-1}$, the plateau of the
inner region shrinks with time as $(t_c-t)^{3/2}$, the power law tail in the
outer region is $\sim x^{2/3}$, and where the Lagrangian velocity, rather than
the Lagrangian acceleration, is constant in time.

\subsection{Time scale separation and isobaric
scenario} \label{timehierarchy}

In general, there are three time scales that characterize the dynamics described
by the nonlinear IGHD equations (\ref{a311}) and (\ref{a322}): the sound travel
time $t_{sound}\sim L/\sqrt{T}$, the cooling time $t_{cooling}\sim
1/(\Lambda\rho\sqrt{T})$, and the inertial time $t_{inertial}\sim L/v$. Here
$\rho$, $T$, and $v$ are typical values of the fields, while $L=L(t)$ is the
characteristic spatial scale of the flow. The evolution of the hydrodynamic
fields can produce time scale separation: a strong inequality between the time
scales. Moreover, the hierarchy of the time scales can be different in different
regions of space. Let us evaluate the time scales for our exact solutions
(\ref{specialp})-(\ref{specialv}). Here there are only two independent time
scales: the sound travel and the cooling time scales. This stems from the fact
that the compressional heating and the inelastic cooling balance each other in
the equation for the pressure, so that $t_{inertial} \sim t_{cooling}$. Consider
the inner region $|x-x_c|\lesssim l(1-t/t_c)^{5/2}$. From the results of the
previous subsection, $\sqrt{T}\sim \sqrt{T_0}(1-t/t_c)$, while $L(t)\sim
l(1-t/t_c)^{5/2}$. As $l/\sqrt{T_0}\sim \tau\sim t_c$, we obtain
\begin{equation}
t_{sound}\sim t_c\left(1-\frac{t}{t_c}\right)^{3/2}.\label{acoustictime}
\end{equation}
Now, using $\rho\sim \rho_0(1-t/t_c)^{-2}$, we find
\begin{equation}
t_{cooling}\sim t_c\left(1-\frac{t}{t_c}\right).\label{coolingtime}
\end{equation}
We observe that, except close to $t_c$, $t_{sound} \sim t_{cooling} \sim t_c$.
However, as the singularity is approached, the sound travel time in the inner
region becomes much shorter than the cooling time. In this situation, the
pressure in the inner region is expected to become constant \cite{meersonRMP} as
our solutions indeed show.

To further elucidate this point, let us estimate the size of the spatial region
at $t=t_c$ such that, within this region, the time scales obey the strong
inequality $t_{sound}\ll t_{cooling}$. Equations~(\ref{acoustictime}) and
(\ref{coolingtime}) show that, at $\sqrt{1-t/t_c}\ll 1$, $t_{sound}\ll
t_{cooling}$ in the inner region. At these times the size $L(t)$ of the
shrinking inner region obeys the inequality $(L/l)^{1/5}\ll 1$. As we saw in the
previous subsection, the shrinking inner region leaves behind stationary
profiles of the fields. Therefore, at $t=t_c$, the local time scales at some
$|x|=x_0\ll l$ can be estimated by their values at those times when $L(t)$
shrinks to the size $x_0$. That is, at $t=t_c$ the time scale separation
$t_{sound}\ll t_{cooling}\sim t_{inertial}$ holds in the region
$(|x-x_c|/l)^{1/5}\ll 1$ which is precisely the region where the pressure is
approximately constant. Therefore, the density blowup, as featured by our exact
solutions, locally follows the isobaric  scenario, previously suggested in the
context of condensation processes developing in gases and plasmas that cool by
their own radiation \cite{meersonRMP,meerson89}. This fact has important
consequences for the theory of clustering that will be explored elsewhere.

We now proceed to the derivation of an additional solution that would allow us
to demonstrate two important features of the developing density singularity: its
locality and its possible coexistence with shock singularities of ordinary gas
dynamics.

\section{A piston moving into a granular gas at rest: a blowup with a shock}
\label{shocks}  The family of exact solutions, describing the finite time
density blowup and reported in Section \ref{analytic}, have a special value of
the total gas mass, $\pi/\mu=(\sqrt{2} \pi \gamma)/{\Lambda}$. Here we relax
this requirement by constructing an exact solution that can have an arbitrarily
large mass. In this solution both the finite-time density blowup and an
``ordinary" shock discontinuity are present.

First, we note that formation of a density singularity in hydrodynamics is a
local process. The set of Eqs.~(\ref{a31})-(\ref{a32}) is hyperbolic and has a
finite speed of propagation of information. Therefore, a finite-time density
blowup, developing at a point with a finite $x$, cannot be affected by a change
in the initial conditions sufficiently far away (provided the velocity is finite
there). In particular, this is true for initial density variations that change
the total mass of the gas, possibly making it infinite. The solution that we are
going to present here illustrates this point, as it has an arbitrary density
distribution sufficiently far from the developing singularity.

The solution also illustrates another point that is absent in the solutions
reported in Section \ref{analytic}: appearance of shocks. At $\Lambda=0$,
Eqs.~(\ref{a31})-(\ref{a32}) become the equations of classical ideal gas
dynamics which produce shocks: initially smooth hydrodynamic fields develop
shock discontinuities in a finite time \cite{Landau,Chorin}. Shocks can also
appear at $\Lambda>0$, the case of our interest. The following argument can be
helpful in elucidating the comparative role of the two types of singularities:
the density blowup and the shock.  Let the initial conditions be fixed. Then, as
$\Lambda$ goes down, the development of a density blow up will be delayed in
time (the delay time becoming infinite as $\Lambda\to 0$). On the other hand,
the time of shock formation must obviously approach a finite limit as
$\Lambda\to 0$. Therefore, for sufficiently small $\Lambda$ the shock formation
will typically  precede the density blowup. In the exact solution that we are
going to present, the shock is present in the solution from the very beginning.

We choose for this solution (an extended version of) a basic setting of ideal
one-dimensional gas dynamics: at $t=0$ a piston starts moving at a constant
speed $v_0$ into an undriven granular gas at rest. Because such a gas has a zero
temperature everywhere, the initial state of the gas is uniquely characterized
by the initial density profile, say $\rho_0(x)$. It is convenient to go over to
the frame moving with the piston, where the piston rests at $x=0$. There one
needs to solve Eqs. (\ref{a311}) and (\ref{a322}) with the initial conditions
$\rho(x, 0)=\rho_0(x)$, $v(x, 0)=-v_0$, $T(x, 0)=0$ and the boundary conditions
$v(x=0, t)= 0$ and $v(x=+\infty, t)= -v_0$. At $t=0$ the gas hits the piston
wall at $x=0$, and a shock forms instantaneously and starts propagating into the
gas. Each of the hydrodynamic fields experiences a discontinuity at the shock
front $x_0(t)$ that obeys $x_0(0)=0$. The solution at $x>x_0(t)$ is of course
$\rho(x, t)=\rho_0(x+v_0 t)$, $v(x,t)=-v_0$ and $T(x,t)=0$, while at $0\leq
x\leq x_0(t)$ non-trivial distributions of the hydrodynamic fields develop. We
show below that a special choice of $\rho_0(x)$ yields a solution that, at
$x<x_0(t)$, is of the same type as that described in Section \ref{analytic}.

\begin{figure}
\includegraphics[width=6.5 cm,clip=]{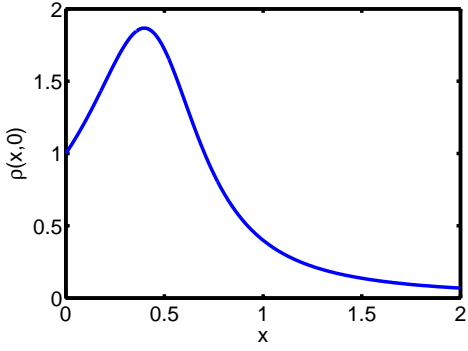}
\caption{The initial density $\rho_0(x)$ (in units of $\rho_0$) versus the
Eulerian coordinate (in units of $l$), see Eq.~ (\ref{indens1}), in the problem
of a piston moving into a granular gas at rest for
$\gamma=2$.}\label{indensprof}
\end{figure}

The jump conditions at the shock front are provided by the same Rankine-Hugoniot
conditions as in the ordinary gas \cite{Landau}. Indeed, these conditions can be
obtained by considering the equations of mass, momentum and energy balance in an
infinitesimal volume $(x_0(t)-\epsilon, x_0(t)+\epsilon)$, in the limit of
$\epsilon\to 0$. While the mass and momentum balances are exactly the same as in
the ordinary gas, the inelastic loss correction to the energy balance is
proportional to $\int_{x_0(t)-\epsilon}^{x_0(t)+\epsilon} \rho^2 T^{3/2}dx$ (see
subsection \ref{energy}), and it vanishes at $\epsilon \to 0$ despite the
discontinuity of the integrand. The Rankine-Hugoniot conditions state that, in
the frame moving with the piston, the mass, momentum and energy fluxes should be
continuous through the shock \cite{Landau}:
\begin{eqnarray}
\rho_2(v_2-\dot {x}_0)&=&-\rho_1(v_0+\dot {x}_0), \label{RG1}\\
\rho_2(v_2-\dot {x}_0)^2&+&\rho_2T_2=\rho_1(v_0+\dot {x}_0)^2, \label{RG2}\\
\rho_2(v_2-\dot {x}_0)^3&+&\frac{2\gamma \rho_2(v_2-\dot
{x}_0)T_2}{\gamma-1}\nonumber
\\&=&-\rho_1(v_0+\dot {x}_0)^3,\label{RG3}
\end{eqnarray}
where the subscripts $1$ and $2$ stand for the upstream and
downstream values of the fields, respectively. These conditions
yield
\begin{eqnarray}&&
\frac{dx_0}{dt}=\frac{(\gamma-1)v_0}{2}+\frac{(\gamma+1)v_2}{2},\label{relation01}\\&&
\rho_2=\frac{(\gamma+1)\rho_1}{\gamma-1},\ \
T_2=\frac{(\gamma-1)\left(v_0+v_2\right)^2}{2}.
\label{relations02}\end{eqnarray} Now let us go over to the Lagrangian
coordinates, see Eq.~(\ref{masscoordinate}). We denote the Lagrangian coordinate
of the shock front by $m_f(t)$:
\begin{equation}
m_f(t)=\lim_{\delta\to 0}\int_0^{x_0(t)-\delta}\rho(x', t)dx',
\label{mf}
\end{equation}
where $\delta>0$. Differentiating Eq.~(\ref{mf}) over $t$ we find
\begin{equation}
\frac{dm_f}{dt}=\rho_2\left(\frac{dx_0}{dt}-v_2\right)=\rho_1\left(\frac{dx_0}{dt}+v_0\right)\,,
\end{equation}
where the second equality holds by virtue of Eq.~(\ref{RG1}). Using this result
together with Eqs.~(\ref{relation01}) and (\ref{relations02}), we obtain
\begin{eqnarray}&&
\frac{dm_f}{dt}=\sqrt{\frac{(\gamma+1) p_2}{2u_1}},\ \
u_2=\frac{\gamma-1}{\gamma+1}\ u_1,\nonumber \\&&
v_2=-v_0+\sqrt{\frac{2p_2u_1}{\gamma+1}}\,m \label{rel}
\end{eqnarray}
where the inverse density $u_{1,2}=1/\rho_{1,2}$ is introduced and $T$ is
expressed via $p$ and $u$. Now we show that it is possible to choose the initial
density profile $\rho_0(m)$ [or, equivalently, $\rho_0(x)$] so that the solution
of Eqs.~(\ref{eqs1}) and (\ref{eqs2}) at $0\leq m\leq m_f(t)$ is a constant
acceleration solution presented in Section \ref{analytic}. In the upstream
region, $m>m_f(t)$, the gas is undisturbed by the shock, and the solution is
$u(m, t)=u_0(m)$, $v(m, t)=-v_0$ and $p(m, t)=0$. In the downstream region
$0<m<m_f(t)$ the hydrodynamic fields at $t>0$ are
\begin{eqnarray}
p(m)&=&2A\cos (\mu m),\label{behindshockp} \\
u(m, t)&=&\left[f(m)-\mu t\sqrt{A\cos (\mu m)}\right]^2,\label{behindshocku} \\
v(m, t)&=&-2\mu \int_0^m f(m')\sqrt{A\cos (\mu m')}dm' \nonumber \\
&+&2A\mu t\sin (\mu m),\label{behindshockv}
\end{eqnarray}
where $f(m)$ is a yet unknown function. By construction, the upstream and
downstream solutions obey the governing equations, and what is left is to impose
the boundary conditions (\ref{rel}) at the shock front. The first condition
becomes
\begin{equation}
\frac{dm_f}{dt}=\sqrt{\frac{A (\gamma+1) \cos[\mu
m_f(t)]}{u_0[m_f(t)]}}.
\end{equation}
Once $u_0(m)$ is known, this equation, together with the initial condition
$m_f(0)=0$, determines the shock coordinate versus time, $m_f=m_f(t)$, via its
inverse function: $t=t_0(m_f)$, where
\begin{equation}
t_0(m)=\int_0^m \sqrt{\frac{u_0(m')}{ A(\gamma+1)\cos(\mu m')}}dm'.
\label{inverse}\end{equation} Using the inverse function $t_0(m)$, we demand the
second and third conditions in Eq.~(\ref{rel}),
\begin{eqnarray}&&
f(m)-\mu t_0(m)\sqrt{A\cos (\mu
m)}=\sqrt{\frac{(\gamma-1)u_0(m)}{\gamma+1}},\
\label{conditions1}\\&& -2\mu \int_0^m f(m')\sqrt{A\cos (\mu m')}dm'
  +2A\mu t_0(m)\sin (\mu m) \nonumber
\\&& =
-v_0+\sqrt{\frac{4Au_0(m)\cos (\mu m)}{\gamma+1}}.
\label{conditions2}
\end{eqnarray}
on the interval $0<m<\pi/(2\mu)$. Using Eqs.~(\ref{inverse}) and
(\ref{conditions1}), we can express $f(m)$ via $u_0(m)$:
\begin{eqnarray}
f(m)&=&\mu \int_0^m\sqrt{\frac{u_0(m')\cos (\mu m)}{(\gamma+1)
\cos(\mu m')}}dm'
\nonumber \\
&+&\sqrt{\frac{(\gamma-1)u_0(m)}{\gamma+1}}. \label{functionf}
\end{eqnarray}
Note that this relation does not include the parameter $A$. We now substitute
Eq.~(\ref{functionf}) into Eq.~(\ref{conditions2}) and arrive at a closed
equation for $u_0(m)$:
\begin{eqnarray}&&
-2\mu^2\int_0^m dm' \cos (\mu m')\int_0^{m'} dm''
  \sqrt{\frac{u_0(m'')}{(\gamma+1)
\cos (\mu m'')}}\nonumber\\&&-2\mu\int_0^m
dm'\sqrt{\frac{(\gamma-1)u_0(m')\cos(\mu m') }{\gamma+1}} \nonumber \\
&&+2\mu \sin(\mu m) \int_0^m \sqrt{\frac{u_0(m')}{(\gamma+1)\cos
(\mu
m')}}dm'\nonumber \\
&&= -\frac{v_0}{\sqrt{A}}+\sqrt{\frac{4u_0(m) \cos (\mu m)}{\gamma+1}}.
\label{cond2}\end{eqnarray} This cumbersome equation is a linear integral
equation for the function $\sqrt{u_0(m)}$, and it is soluble analytically.
Changing the order of integration in the first term of the equation, we rewrite
the first term as
\begin{eqnarray}&&
-2\mu^2\int_0^m dm''\sqrt{\frac{u_0(m'')}{(\gamma+1) \cos(\mu
m'')}}\int_{m''}^m\cos (\mu m')dm' \nonumber\\&& =-2 \mu \sin(\mu m)\int_0^m
dm''\sqrt{\frac{u_0(m'')}{(\gamma+1) \cos (\mu m'')}} \nonumber\\&&
+2\mu\int_0^m dm'' \sin(\mu m'')\sqrt{\frac{u_0(m'')}{(\gamma+1) \cos (\mu
m'')}}. \nonumber\end{eqnarray} This brings a partial cancelation of terms in
Eq.~(\ref{cond2}), and we obtain a simpler equation
\begin{eqnarray}
&&2\mu\int_0^m dm' \sin(\mu m')\sqrt{\frac{u_0(m')}{(\gamma+1)
\cos(\mu m')}}
\nonumber \\
&&-2\mu\sqrt{\frac{(\gamma-1)}{\gamma+1}} \int_0^m
dm'\sqrt{u_0(m')\cos (\mu
m')}\nonumber \\
&&= -\frac{v_0}{\sqrt{A}}+2\sqrt{\frac{u_0(m) \cos (\mu m)}{\gamma+1}}.
\nonumber\end{eqnarray} Now let us introduce an auxiliary function
$$g(m)=2\sqrt{\frac{u_0(m)
\cos(\mu m)}{\gamma+1}}$$ that obeys, on the interval
$0<m<\pi/(2\mu)$,  a linear integral equation:
\begin{equation}
\mu \int_0^{m}g(m')\left[\tan (\mu m')-\sqrt{\gamma-1}\right]dm' =
-\frac{v_0}{\sqrt{ A}}+g(m) \label{init}
\end{equation}
The solution for $g(m)$ is elementary:
\begin{equation}
g(m)=\frac{v_0\exp\left(-\sqrt{\gamma-1}\,\mu m\right)}{\sqrt{A} \cos (\mu
m)}\,,
\end{equation}
so the result for $u_0(m)$ is
\begin{equation}
u_0(m)=\frac{v_0^2(\gamma+1)\exp\left(-2\sqrt{\gamma-1}\,\mu m\right)}
{4A\cos^3(\mu m)}\,. \label{initialu}\end{equation} To complete the formal
construction of the solution, we present the initial gas density in the Eulerian
coordinates, $\rho_0(x)$, in a parametric form:
\begin{eqnarray}&&
\rho_0(m)=\rho_0 \cos^3 (\mu m) \exp\left(2\sqrt{\gamma-1}\,\mu
m\right),\nonumber
\\&& x=\int_0^m \frac{dm'}{\rho_0(m')}\,,
\label{indens1}\end{eqnarray} where $\rho_0=4 A/[(\gamma+1) v_0^2]$. The graph
of $\rho_0(x)$ is shown in Fig.~\ref{indensprof}.

\begin{figure}
\includegraphics[width=8.5 cm,clip=]{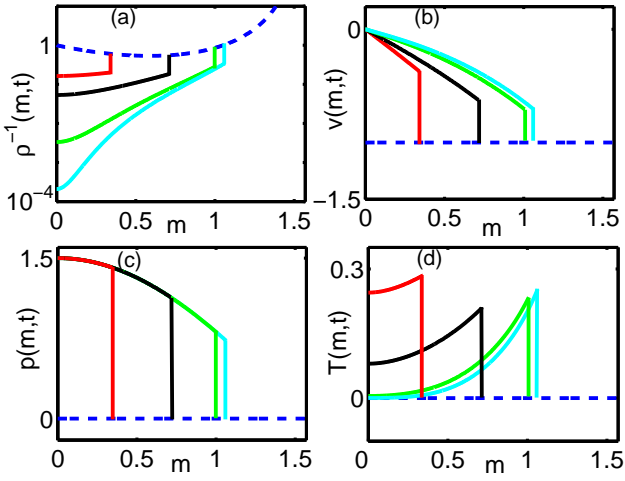}
\caption{(Color online) The exact solution~(\ref{pistonfinal}), in the
Lagrangian coordinates, of the problem of a piston moving into a granular gas at
rest. Shown are the inverse density (a), velocity (b), pressure (c) and
temperature (d) at rescaled times (from left to right) $t v_0/l=0.2$, $0.4$,
$0.6$ and $0.65$ as a function of $m$ measured in units of $1/\mu$ (the solid
lines). The dashed lines are the initial conditions: the initial density from
Eq.~(\ref{indens1}), the constant (rescaled) velocity $-1$, and zero temperature
and pressure.  The density, velocity, pressure and temperature are rescaled to
$\rho_0$, $v_0$, $\rho_0 v_0^2$ and $v_0^2$, respectively, and $\gamma=2$. The
density blows up at the piston, $m=0$, at the rescaled time $t_cv_0/l = t_*
v_0/l =2/3$. The rescaled Lagrangian coordinate of the shock at the blowup time
is $1.08031\dots$.}\label{figpist}
\end{figure}

Now we use Eq.~(\ref{initialu}) to calculate the inverse function $t=t_0(m_f)$
from Eq.~(\ref{inverse}) that determines the shock motion law in the Lagrangian
coordinates, $m_f=m_f(t)$:
\begin{equation}
t_0(m)=t_*\Phi(\mu m),\ \ \Phi(z)= \int_0^z
\frac{\exp\left(-\sqrt{\gamma-1}\,z'\right)dz'} {\cos^2 z'}\,,
\nonumber\end{equation} where we have introduced the characteristic inelastic
cooling time scale $t_*=2l/[(\gamma+1)v_0]$, and $l=1/(\mu\rho_0)$ is the
inelastic cooling length scale. As $\Phi(z)$ diverges at $z=\pi/2$,  $m_f(t)$
satisfies, at any finite time, the double inequality $0< m_f(t)< \pi/(2\mu)$. We
reiterate that, in the upstream region $m>m_f(t)$, the gas is unperturbed by the
shock. The final form of the solution in the downstream region, $0\leq
m<m_f(t)$,  is
\begin{eqnarray}
p(m, t)&=&\frac{(\gamma+1)\rho_0v_0^2\cos(\mu m)}{2},\nonumber \\
\rho(m, t)&=&\frac{\rho_0(\gamma+1)}{\cos(\mu m)\left[\Phi(\mu
m)+\sqrt{\gamma-1}\Phi'(\mu m)-t/t_*\right]^2},\nonumber \\
v(m, t)&=&-v_0\int_0^{\mu m} dz \cos z
\left[\Phi(z)+\sqrt{\gamma-1}\Phi'(z)\right] \nonumber \\
&+&v_0 (t/t_*) \sin(\mu m)\,.\label{pistonfinal}
\end{eqnarray}
This solution is shown in Fig.~\ref{figpist}. Note that, at fixed $\rho_0$ and
$v_0$, the characteristic spatial and temporal scales of the solution behave as
$1/\Lambda$. For this solution, the gas density blows up in a finite time at the
point $m$ where the function $\Phi(\mu m)+\sqrt{\gamma-1}\Phi'(\mu m)$ has its
minimum. One can easily see that, for $\gamma\leq 2$, this function is monotone
increasing with $m$. As a consequence, the density blows up at the piston (that
is, at $m=0$), and this happens at $t=t_c=\sqrt{\gamma-1}\,t_*$. At this time,
the shock front location $m_*\equiv m_f(t_*)$ is described by the relation
$\Phi(\mu m_*)=\sqrt{\gamma-1}$ which yields $\mu m_* =1.08031\dots$ for
$\gamma=2$ (in 2d) and $0.87915 \dots$ for $\gamma=5/3$ (in 3d). That is, by the
time $t=t_c$ when the density blows up at the piston, the shock has traveled
only a finite distance (both in the $m$-, and in the $x$-space) from the piston.
It is obvious, therefore, that our solution allows an \textit{arbitrary}
modification of the density profile $\rho_0(x)$ at sufficiently large $x$ that
are unreachable for the shock. This clearly shows that initial states with an
arbitrary large mass of gas can develop a finite-time density blowup.

\section{Numerical solutions}
\label{numerics}

We confirmed the exact solutions, presented in Figs.~1-4 and 6,  by solving
numerically the IGHD equations (\ref{eqs1}) and (\ref{eqs2}) in the Lagrangian
coordinates. In each case we used the initial and boundary conditions, provided
by the exact solutions themselves. We employed the classical artificial
viscosity, staggered grid scheme of von Neumann and Richtmyer \cite{code}. The
number of cells (grid points) used varied between 1000 and 2000 and showed
numerical convergence until close to the singularity. As the solution developed
a density blowup, the simulation had to be terminated at a very high but finite
maximum density: usually at about $10^7 \rho_0$, where $\rho_0$ is the initial
density. We also enforced the simulations to stop when the density jump between
two adjacent cells exceeded a prescribed value, usually 50 percent. (When such
jumps develop near the singularity, the accuracy of the simulation degrades and
can be restored only by a rezoning algorithm that was not employed.) One example
of the numerical solution is shown in Fig.~\ref{fig1} alongside with the
analytical solution, and very good agreement is observed. Very good agreement
was also obtained for the solutions shown in Figs.~3, 4 and 6 (not shown). The
numerical solutions inevitably add some effective noise to the system because of
the spatial and temporal discretization and round-off errors. The fact that the
analytical solutions are accurately reproduced numerically confirms their
stability with respect to small one-dimensional perturbations.

\begin{figure}
\includegraphics[width=8.7 cm,clip=]{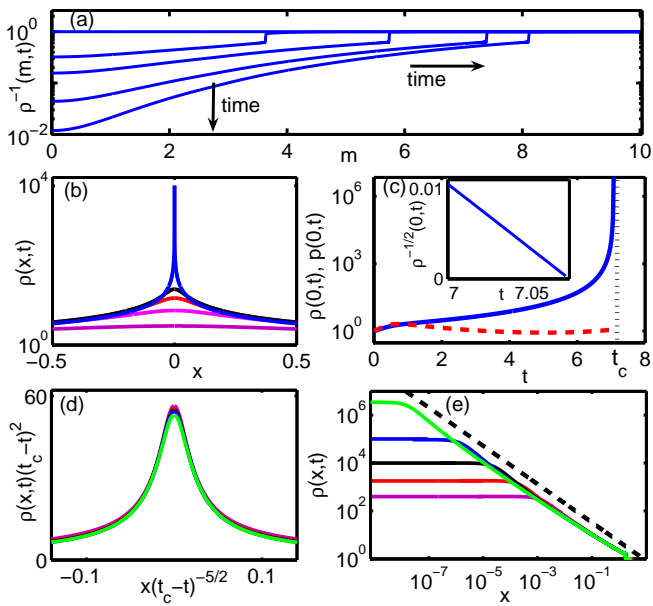}
\caption{(Color online) An example of numerical solution of Eqs.~(\ref{eqs1})
and~(\ref{eqs2}). The initial conditions are $\rho(m,0)=\rho_0$, $T(m,0)= T_0$
and $v(m,0)=-v_0\tanh\left[m/(\rho_0 \lambda)\right]$. (a) The inverse density
profiles at rescaled times $0$, $2.06$, $3.76$, $5.36$ and $6.26$. (b) The
density $\rho(x,t)$ in the logarithmic scale at times $t=2$, $4$, $5$, $5.5$ and
$7$ (from the bottom to the top, respectively) as a function of $x$. (c) The
time history of the density (the solid line) and pressure (the dashed line),
both in the logarithmic scale, at the blowup point $x=0$. The inset depicts
$\rho^{-1/2}(x=0,t)$ in the logarithmic scale versus time in a vicinity of
$t=t_c$. (d) The rescaled density $\rho(x,t)(t_c-t)^2$ as a function of the
rescaled coordinate $x(t_c-t)^{-5/2}$ at times $t=6.7$, $6.9$, $7$, $7.05$ and
$7.07$ (from the top to the bottom, respectively) in a vicinity of the
singularity. The profiles at different times almost coincide demonstrating the
local self-similarity of the blowup.(e) The log-log plot of the density
$\rho(x,t)$ versus $x$ (solid lines) for the same times as in (c) (from the
bottom to the top, respectively). The dashed line shows a power law $\rho\sim
x^{-4/5}$, to guide the eye. The Lagrangian and Eulerian coordinates, time and
the hydrodynamic fields are made dimensionless, as explained in
Section~\ref{numerics}. The parameters are $\tilde{\Lambda}= \rho_0 \Lambda
\lambda=0.5,\,v_0=\sqrt{T_0}$ and $\gamma=2$.} \label{tanh}
\end{figure}

We also performed extensive numerical simulations with Eqs.~(\ref{eqs1}) and
(\ref{eqs2}) for a variety of initial and boundary conditions that did not
correspond to any of the exact solutions.  As already briefly reported in
Ref.~\cite{Fouxon1}, these simulations show that, for generic initial
conditions, a finite-time density blowup always occurs. Remarkably, the
numerical solutions exhibit, close to the singularity, the same local scaling
behaviors in space and in time as those exhibited by our exact solutions and
presented in subsection \ref{local}. One series of simulations dealt with a
symmetric inflow of an initially uniform gas, $\rho(x,0)=\rho_0$ and
$T(x,0)=T_0$, from ``infinity": $v(x,0)=-v_0 \tanh(x/\lambda)$. Here it is
convenient to rescale the $x$-coordinate by $\lambda$, time by
$\lambda/\sqrt{T_0}$, the density by $\rho_0$, the velocity by $\sqrt{T_0}$, the
temperature by $T_0$, the pressure by $\rho_0 T_0$ and the Lagrangian mass
coordinate $m$ by $\rho_0 \lambda$. After this rescaling the governing equations
Eqs.~(\ref{eqs1}) and~(\ref{eqs2}) keep their form, except that $\Lambda$
becomes rescaled: $\tilde{\Lambda}= \rho_0 \Lambda \lambda$. The rescaled
initial conditions become $\rho(x,0)=p(x,0)=1$ and
$v(x,0)=-(v_0/\sqrt{T_0})\tanh (x)$. The numerical solutions were obtained on
the (rescaled) interval $|x|<10$ that corresponds, at $t=0$, to the (rescaled)
Lagrangian interval $|m|<10$. The boundary conditions are $v(x=\pm 10,
t)=v(x=\pm 10, 0)=\mp (v_0/\sqrt{T_0}) \tanh 10$. Because of the symmetry of the
problem with respect to $x=0$ we actually solved it on the half-interval
$0<x<10$, replacing the boundary condition at $x=-10$ by the condition
$v(0,t)=0$.

Here we present one typical example of such a simulation, briefly mentioned in
Ref.~\cite{Fouxon1}. The parameters are $\tilde{\Lambda}=0.5,\,v_0=\sqrt{T_0}$
and $\gamma=2$. The (one half of the) simulated flow is shown in
Fig.~\ref{tanh}. Panel (a) provides a general view of (one half of) the system.
The gas inflow creates a compression in the central region. A compression wave
propagates outwards and steepens. This steepening would lead to a wave breaking
singularity, but the numerical scheme resolves this singularity, by means of the
artificial viscosity, as a shock wave that continues propagating outward. At the
same time the gas density at the origin continues growing and ultimately blows
up. The density growth in a vicinity of the origin is presented in panel (b).
Panel (c) focuses on the density and pressure history at the origin. While the
density blows up at $t=t_c$, the pressure hardly varies there so the isobaric
scenario holds. The inset of panel (c) verifies that, close to $t_c$, the
density blowup proceeds as $(t_c-t)^{-2}$. Panels (d) and (e) add more tests of
the spatial and temporal scaling behavior near the singularity. Panel (d) shows
a plot of the rescaled density $\rho(x,t)(t_c-t)^2$ versus the rescaled
coordinate $x(t_c-t)^{-5/2}$. The collapse of the curves at different times
proves the local self-similarity (note that the gas density at the origin
varies, for these times, by four orders of magnitude). Finally, panel (e)
verifies the presence of the inner and outer regions, described by our theory,
see subsection~\ref{local}. The density plateau, whose size shrinks as $\sim
(t_c-t)^{5/2}$, represents the inner region, while in the outer region a
time-independent density profile forms with a power-law behavior $\rho \sim
x^{-4/5}$ as predicted by our exact solutions.

The same universal features of the singularity were also observed when starting
from small-amplitude sinusoidal density or velocity perturbations around a
homogeneous state.

\section{Non-ideal effects near the singularity} \label{regularization}

Having found the exact nonlinear solutions of the IGHD equations (\ref{a311})
and (\ref{a322}), we are in the position to test the assumptions behind these
equations and establish the domain of validity of the solutions (as accurate
leading-order descriptions) in the presence of ``non-ideal" effects. First, the
validity of the Navier-Stokes hydrodynamics,
Eqs.~(\ref{hydrodynamics1})-(\ref{hydrodynamics3}), demands that the Knudsen
number $l_{free}/L$ remain much smaller than unity. The smallest hydrodynamic
length scale of the solution can be estimated as $L(t)\sim l (1-t/t_c)^{5/2}$,
see subsections \ref{local} and \ref{timehierarchy}. The mean free path
$l_{free}\sim 1/(\rho\sigma^{d-1})\sim (1-t/t_c)^2/(\rho_0\sigma^{d-1})$. As
$l\sim 1/(\Lambda\rho_0)$ and $\Lambda\sim (1-r)\sigma^{d-1}$, we obtain
\begin{equation}
\frac{l_{free}}{L}\sim \frac{1-r}{\sqrt{1-t/t_c}}\,.\label{freepath}
\end{equation}
At $t \lesssim t_c$ the  Knudsen number is small by the assumption $1-r\ll 1$.
However, as $t$ approaches $t_c$, the Knudsen number grows indefinitely which
invalidates the hydrodynamics.  We shall see, however, that one of the
assumptions of the IGHD breaks down even earlier.

Now consider the ratio of the viscous stress term to the pressure gradient term
in the momentum equation (\ref{hydrodynamics2}). This ratio can be estimated as
$(\nu_0\sqrt{T}\partial_x v)/(\rho T)=\nu_0\partial_mv/\sqrt{T}$. We first
estimate all the ratios in the inner region, see subsection \ref{local}. As both
$\partial_m v$ and $\sqrt{T}$ vanish linearly as $t \to t_c$, their ratio
depends on time only weakly, and can be estimated by its value at $t=0$. As
$|\partial_m v(m, 0)|\sim \Lambda\sqrt{T}$ (see subsection \ref{Lagrangian}), we
find
\begin{equation}
\left|\frac{\nu_0\partial_x(\sqrt{T}\partial_x v)}{\partial_x(\rho
T)}\right|\sim 1-r\,,
\end{equation}
that is the viscous stress is always negligible in our solutions as long as $1-r
\ll 1$.

The same estimate (up to the sign) holds for the ratio of the viscous heating
term to the compressional heating term in Eq.~(\ref{hydrodynamics3}):
\begin{equation}
\left|\frac{\nu_0(\gamma-1)\sqrt{T}(\partial_x v)^2/\rho}{(\gamma-1)T\partial_x
v}\right|\sim 1-r\ll 1\,.
\end{equation}
In contrast to the viscous terms, the heat conduction term in
Eq.~(\ref{hydrodynamics3}), which is initially small in our solutions, does
become important near the singularity. We estimate the ratio of the heat
conduction term to the inelastic cooling term as follows:
\begin{equation}
\left|\frac{\kappa_0\partial_x(\sqrt{T}\partial_x T)/\rho}{\Lambda \rho
T^{3/2}}\right|\sim \frac{\kappa_0}{\Lambda \rho^2 L^2}  \sim
\frac{1-r}{1-t/t_c}\,.
\end{equation}
The same estimate is obtained for the ratio of the heat conduction term to the
compressional heating term. This ratio becomes of order unity at $1-t/t_c\sim
1-r \ll 1$. At this time $l_{free}/L\sim \sqrt{1-r}\ll 1$, see
Eq.~(\ref{freepath}), so the IGHD equations break down  while the full
Navier-Stokes hydrodynamics, Eqs.~(\ref{hydrodynamics1})-(\ref{hydrodynamics3}),
is still valid.

In its turn, the dilute gas assumption, $\rho \sigma^d \ll 1$, breaks down, and
excluded particle volume effects become important, at $1-t/t_c\sim
\sqrt{\rho_0\sigma^d} \ll 1$.  The relative role of the heat conduction and
excluded particle volume effects in the breakdown of our analytic solutions near
the attempted singularity is determined by the competition between the small
parameters $\sqrt{\rho_0\sigma^d}$  and $1-r$. When
$\sqrt{\rho_0\sigma^d}\gtrsim 1-r$, our solutions remain valid until the density
becomes comparable with the (fraction of) close packing density. This happens at
$1-t/t_c\sim \sqrt{\rho\sigma^d}$, so $L_{valid}\sim l(\rho_0\sigma^d)^{5/4}$.
At moderate densities one can use, in numerical solutions, a half-empiric
equation of state due to Carnahan and Starling \cite{Carnahan}, and half-empiric
transport coefficients obtained for granular gases \cite{Jenkins} in the spirit
of Enskog kinetic theory \cite{Resibois}.

When $\sqrt{\rho_0\sigma^d}\ll 1-r$ the heat conduction becomes important when
the gas is still dilute. This happens at time $1-t/t_c\sim 1-r$ so that
$L_{valid}\sim l(1-r)^{5/2}$. As long as $\sqrt{1-r}\ll 1$, the Knudsen number
is still small at that time, and the complete Navier-Stokes hydrodynamics is
still applicable.

Therefore, as the blowup time $t_c$ is approached, either the heat conduction,
or the excluded particle volume effects become important and invalidate our
theory in the inner region. In this sense, our solutions describe an
intermediate asymptotic regime of formation of a dense cluster of particles.
Importantly, in  the outer region our solutions remain valid until much later
times. Indeed, as the inner region shrinks, it leaves behind stationary profiles
of the fields, see subsection \ref{local}. The ratios of the different terms
governing the IGHD validity are given, at some $x$ from the outer region, by the
corresponding ratios in the inner region estimated at the time when $L(t) \sim
|x-x_c|$. As a result, at some time close to $t_c$ (see above), our solutions
break down in the inner region which size at that time is $L_{valid}$, while in
the outer region, $|x-x_c|\gg L_{valid}$, our solutions continue to hold.

The fact that the IGHD description breaks down only locally, in a small region
of space, can be exploited for derivation of an \textit{effective} description
of the clustering dynamics beyond the singularity time. In this description
close-packed clusters appear as finite-mass point-like singularities of the
density \cite{Fouxon1}. This effective description is similar in spirit to the
description of shocks (that actually have finite widths) as discontinuities in
ordinary ideal gas dynamics.

\section{Summary} \label{summary}

Let us briefly summarize the main results of this work. We introduced ``ideal
granular hydrodynamics" (IGHD) equations that provide a simple but informative
description of non-stationary large-scale flows of dilute granular gases with
nearly elastic collisions between the particles. We employed the IGHD equations
to investigate analytically and numerically the paradigmatic phenomenon of
particle clustering in freely cooling granular gases. We believe that the IGHD
will provide a useful framework for a host of other problems involving
large-scale flows of \textit{driven} granular gases.

We focused on a one-dimensional granular hydrodynamic flow, characteristic of an
idealized channel geometry, and derived a family of exact nonlinear and
non-stationary analytic solutions of the IGHD that describe a finite-time blowup
of the gas density. The derivation was made possible by a transformation of the
hydrodynamic equations to Lagrangian coordinates. The exact solutions are
characterized by a constant in time accelerations in the Lagrangian frame, and
they are not self-similar. We investigated the local structure of the flow near
the singularity and determined local spatial and temporal scaling laws for the
hydrodynamic quantities in question. We also found an instructive soluble case
of the problem of a piston entering, at a constant speed, a granular gas at
rest. Here a density singularity, developing at the piston, coexists with a
shock wave that is located elsewhere. Besides demonstrating the presence of the
two different types of singularities in the same system, this solution allows an
arbitrary density profile at large distances, showing that the density blowup is
a local process. In all of these solutions the developing singularity obeys
locally the isobaric scenario \cite{meersonRMP,meerson89}: the gas pressure
remains approximately uniform in space and constant in time in a close vicinity
of the blowup point. This finding has important consequences for the theory of
clustering. Indeed, by imposing, in the zeroth order of theory, the isobaricity
condition on the hydrodynamic equations, one arrives at a powerful reduced
description of the nonlinear clustering process. This description is valid for a
much broader class of initial conditions than those giving rise to the exact
solutions reported in the present work. The corresponding results will be
presented elsewhere.

Our numerical solutions of the IGHD equations accurately reproduce the analytic
solutions, thus confirming their stability with respect to small one-dimensional
perturbations. Furthermore, numerical simulations with a variety of initial and
boundary conditions showed that that the local scaling laws near the singularity
are universal, that is independent of details of the initial and boundary
conditions.

We also analyzed additional physical effects, neglected in IGHD, which become
important in a narrow spatial region near the attempted singularity. Depending
on the parameters, either excluded particle volume effects, or heat conduction
invalidate our solutions there. In the former case, the density growth should
directly cross over to the formation of close-packed clusters of particles. In
the latter case, the final outcome of the nonlinear density growth is yet
unknown. The future work should find out whether the heat conduction arrests the
density blowup or only modifies the singularity law. In any case, our exact
solutions can be viewed as instructive intermediate asymptotics, describing a
broad class of strongly nonlinear clustering flows of granular gases. No less
important, as breakdown of these solutions occurs only in the narrow regions
around the density peaks, it is possible to continue the ideal solutions
\textit{beyond} the singularity by introducing into the theory finite-mass
point-like singularities of the density. This procedure yields a simple
effective description of granular gases with embedded close-packed clusters, in
much the same way as the ideal gas dynamics describes a gas flow with shock
discontinuities \cite{Fouxon1}.

Very recently, one of our exact solutions has been successfully tested against
MD simulations of the dynamics of a very dilute gas of nearly elastically
colliding identical hard disks in a very long and narrow 2d channel. The results
will be presented elsewhere.

Finally, we note that the power law of the density blowup near the singularity,
$\sim (t_c-t)^{-2}$, is the same as the one recently found in a different
setting: that of an initially thermalized granular gas, freely cooling and
collapsing under gravity \cite{Volfson}. In each of the two problems the
momentum equation is ``fast", while the energy equation is ``slow" in a small
vicinity of the collapse. This suggests common physics behind the two collapse
phenomena and demands further investigation.

\begin{acknowledgments}
We are grateful to Lev S. Tsimring and Eli Waxman for useful discussions. This
work was supported by the Israel Science Foundation (grant No. 107/05) and by
the German-Israel Foundation for Scientific Research and Development (Grant
I-795-166.10/2003).
\end{acknowledgments}

\end{document}